\documentclass[useAMS,usenatbib,usegraphicx]{mn2e}
\usepackage{amsmath}
\usepackage{aas_macros}
\usepackage{multirow}
\usepackage{url}
\usepackage{tablefootnote}
\usepackage{xcolor}
\usepackage{hyperref}	
\usepackage{lscape}
\hypersetup{colorlinks=true,linkcolor=blue,citecolor=blue,filecolor=blue,urlcolor=blue}
\bibpunct{(}{)}{,}{a}{}{,} 
\newcommand{\lt}{<}
\newcommand{\gt}{>}

\newcommand{\lsun}{L$_{\sun}$}
\newcommand{\msun}{M$_{\sun}$}

\newcommand{\ml}{$\alpha_\mathrm{ML}$}
\newcommand{\teff}{$T_\mathrm{eff}$}
\newcommand{\code}{\textsc{Alice}}

\newcommand{\krad}{$\kappa_\mathrm{rad.}$}
\newcommand{\kcon}{$\kappa_{\rm{con.}}$}

\newcommand{\ids}{\color{blue}}

\title[New TRGB models]{New calibrated models for the TRGB luminosity and a thorough analysis of theoretical uncertainties}

\author[Ippocratis D. Saltas \& Emanuele Tognelli]{
Ippocratis D. Saltas$^{1}$\thanks{E-mail: saltas@fzu.cz}
Emanuele Tognelli,$^{2}$\thanks{E-mail: tognelli@fzu.cz}
\\
$^1$ CEICO, Institute of Physics of the Czech Academy of Sciences, Na Slovance 2, 182 21 Praha 8, Czechia\\
}

\begin{document}
\date{Accepted 20xx January xx. Received 20xx January xx; in original form 20xx January xx}
\pagerange{\pageref{firstpage}--\pageref{lastpage}} 
\pubyear{2022}
\maketitle
\label{firstpage}
%
\begin{abstract}
The luminosity of the Tip of the Red Giant Branch (TRGB) is instrumental for the construction of the distance ladder, and its accurate modelling is key for determining the local Hubble parameter. In this work, we present an extensive quantitative analysis of the TRGB luminosity, accounting for virtually all input physics that affect it: chemical composition, opacity, diffusion, nuclear reaction rates, electron screening, neutrinos, convection efficiency, boundary conditions and mass loss, amongst others. 

Our analysis is based on a newly produced grid of $\sim 3\times 10^6$ TRGB models, evolved from pre-main sequence up to the helium ignition at the TRGB, and covering a wide range of metallicity (Z = 0.0001-0.02) and initial mass (M = 0.8-1.4~\msun). Through a Monte-Carlo analysis, we study the systematic variation of the TRGB luminosity due to the combined effect of all above input physics, and show that a maximum theoretical uncertainty of about $1.6 \%$ is still present on the current generation of models, dominated by systematics of radiative opacity. Results are also provided in several photometric bands.

As a by-product of our analysis, we demonstrate robust evidence for the linear response of the Tip luminosity to individual changes of input physics, which can significantly simplify future analyses. A comparison of our results with other stellar evolution codes shows excellent agreement, while our grid of models is available upon reasonable requests. 

\end{abstract}
\begin{keywords}
Stellar modelling, red giants, TRGB luminosity, Hubble parameter
\end{keywords}
\maketitle
\section{Introduction}
The construction of the distance ladder in the Universe is instrumental in astrophysics and cosmology. Past years, the  accurate inference of distances towards robust determinations of the local value of the Hubble parameter has become a necessity due to the puzzling issue of Hubble tension (see e.g \cite{Riess_2021}, \cite{Freedman_Hubble} ). The latter, translates to the $~ 5\sigma$ discrepancy between the Hubble parameter measured from the CMB ($H_0 = 67.4 \pm 0.5 \; \rm{km} \, \rm{sec}^{-1} \, \rm{Mpc}^{-1}$, \citet{refId0}) and the one measured locally through Cepheid variable stars ($H_0 = 73.2 \pm 1.3 \; \rm{km} \, \rm{sec}^{-1} \, \rm{Mpc}^{-1}$, \cite{Riess_2021}). Interestingly enough, inferences based on the Tip luminosity of red giant stars yields a value much closer to the cosmological one ($H_0 = 69.6 \pm 1.9 \; \rm{km} \, \rm{sec}^{-1} \, \rm{Mpc}^{-1}$, \citet{Freedman_2020}), implying a possible issue with Cepheid measurements. As of now, there is no conclusion as to whether new physics needs to be introduced, or whether it is merely an issue associated to observations. Proposed theoretical resolutions to the problem tend to modify physics either at cosmological or at stellar scales through the introduction of new degrees of freedom (see e.g \cite{Ballesteros:2020sik, Braglia:2020iik, Desmond:2019ygn,Desmond:2020wep, Sabla:2022xzj}). It is understood that the key aspect for a resolution is the need for a robust modelling of astrophysical standard candles and a thorough understanding of the associated systematics. 

Over the years, the {\it Tip Luminosity of the Red Giant Branch} (TRGB) has established itself as a highly reliable standard candle, able to compete with traditional methods such as Cepheid variable stars \citep[][]{Salaris_Cassisi,Bildsten_2017,serenelli17, Freedman_2020}. What makes stars at the TRGB stage act as standard candles is the fact that their maximum (Tip) luminosity, reached when helium starts to ignite in their core, is to a large extend universal, i.e it tends to be approximately the same across different evolutionary tracks. In general, their evolution and dynamics can be modelled in a more accurate manner compared to Cepheids. 

Despite the universal features of the TRGB luminosity, the story does not end there. Uncertainties associated with input physics, such as modelling of opacity or nuclear reaction rates, still leave a sizable residual uncertainty on the TRGB luminosity. Such systematics need be under control, if accurate and precise determinations of distances and the local $H_0$ are to be established.
Overall, uncertainties in the modelling of the TRGB luminosity typically relate to the chemical composition (e.g metallicity), the modelling of opacities (radiative or conductive), nuclear reaction rates from the main sequence stage to the TRGB, neutrino emission, or environmental effects such as mass loss. 

The theoretical uncertainties of the Tip luminosity has been discussed in several previous works. For the most recent we refer to \citet{viaux13}, \cite{valle13a,valle13b}, \citet{serenelli17} \cite{straniero20} and \cite{cassisi21}. Some of these works focused on the effect on the predicted Tip luminosity \citep[e.g.][]{valle13a,valle13b}, while others on Tip magnitudes \citep[][]{viaux13,serenelli10,straniero20,cassisi21}. In particular, \citet{viaux13} studied in detail the impact of each individual quantity that could potentially affect the Tip magnitude, but they considered a tight range of mass (0.8-0.85\msun) and metallicity (Z=0.001-0.0017). \citet{valle13a,valle13b} performed a similar analysis adopting 6 metallicities (Z=0.0001-0.02) for a fixed mass (M=0.9\msun), but the effect on the magnitudes was not studied. However, they studied both the effect of each individual perturbation as well as the significance of non-linear effects on the Tip luminosity allowing for a simultaneous, random perturbation of input physics. More recently, \citet{serenelli17} analysed the errors on the Tip luminosity and magnitudes for a wider range of masses 0.8-1.4\msun{} and metallicity Z=0.0001-0.04, without analysing the effect of overshooting, mass loss and boundary conditions. Our work extends all previous works in the literature, and presents with a complete analysis of input physics that affect the Tip luminosity/magnitude, within a wide range of metallicity, helium abundance, mixing length and mass. 

We first construct a new set of TRGB models, consistently evolved from the pre-main sequence through the red giant branch up to their maximum luminosity at the TRGB stage, and covering virtually all different input physics.
The end result is a grid of Tip models, with a prediction of the Tip luminosity, total radius and effective temperature. Whereas the inference of distances predominantly relies on the Tip luminosity, observables such as the radius and effective temperature are important for complementary studies such as asteroseismic ones. 

Our results show that the dominant uncertainty on TRGB luminosity comes from the modelling of radiative opacities, and translate to an uncertainty on the Tip luminosity of {\ids $1.6 \%$}. 

In summary, the {\it goals of this work} are as follows:
\begin{itemize}
\item[] 1. The presentation of a new set of $\sim 3\times 10^6$ 
TRGB models, evolved from pre-main sequence up to the helium ignition at the TRGB, based on a grid of virtually all relevant input physics which affect the Tip luminosity, and within the sufficiently a wide range of metallicity ($Z=0.0001-0.02$) and initial mass range ($M = 0.8 - 1.4 M_{\odot}$). \\

\item[] 2. A detailed exposition of the qualitative and quantitative effect of each individual input physics on the Tip luminosity, supported by the results of our simulations. Our results are also presented in several photometric bands. To aid future asteroseismic analyses, we also quote some results on the respective effect on radius and effective temperature at Tip luminosity. We also discuss in detail the accuracy of our numerical procedure. \\

\item[] 3. A Monte Carlo analysis where the simultaneous effect of input physics on the Tip luminosity is computed, and its comparison with the simpler analysis where individual uncertainties are added linearly. This will provide a robust verification of the previously claimed linear response of TRGB models under small variations of input physics, and will significantly simplify future computations in this context. \\
\end{itemize}

Our paper is split as follows. In Section \ref{sec:theory} we set the scene through  some intuitive theoretical arguments, and then in Section \ref{sec:Assumptions} we proceed explaining our assumptions regarding the main input physics and numerical modelling. In Section \ref{sec:uncertainties} we detail our results for the uncertainty of the Tip luminosity from the modelling of each individual input physics parameter, while Section \ref{sec:full_analysis} presents our Monte-Carlo analysis and its comparison with the previous linear analysis. The relevant systematic uncertainties on photometric bands are discussed in Section \ref{sec:magnitudes}. Finally, a summary of our results is presented in Section \ref{sec:full_analysis}.

\section{Some basic theoretical considerations} 
\label{sec:theory}
When the central hydrogen at the stellar center is exhausted, the star exits the Main Sequence (MS) phase and enters its ascent to the Red Giant Branch (RGB). This ascent is predominantly characterised by the burning of hydrogen in a shell off the star's center driven by the CNO nuclear cycle. The thickness of the shell quickly reduces, whereas at the same time a compact and degenerate helium core starts to develop. This process leads to a gradual increase of the helium core mass, however, its radius does not change drastically. Throughout this phase, the star is characterised by an increase in its luminosity accompanied by a decrease in effective temperature, eventually reaching a peak luminosity known as the Tip of the Red Giant (TRGB). 

After the formation of a sufficiently big and degenerate helium core in the center, alongside a thin hydrogen burning shell surrounding it, the thermal properties of the star are predominantly governed by the properties of the degenerate core. Although an accurate description requires numerical simulations, we can still gain intuition on the basic scaling properties of quantities such as luminosity and temperature from homology arguments. In this regime, temperature, luminosity, density and pressure can be expressed through scaling relations as
$L \sim M_{\rm{c}}^{a_1} \cdot R_{\rm{c}}^{a_2}, \; \; T \sim  M_{\rm{c}}^{b_1} \cdot R_{\rm{c}}^{b_2}$
and similar relations hold for density and pressure, where $M_{\rm{c}}$ is the mass of helium core. The temperature here refers to that of the hydrogen burning shell and subsequently, to the temperature of the core's surface. The coefficients $a_j, b_j$ are constants which, however, depend on the assumptions for energy generation in the surrounding shell and the opacity. It is convenient to write the aforementioned relations in differential form as \citep[see e.g][]{kippenhahn90}
\begin{align}
& \frac{d \ln L}{d \ln M_{\rm{c}}} = a_{1} + a_{2} \frac{d \ln R_{\rm{c}} }{d \ln M_{\rm{c}}} \nonumber \\
& \frac{d \ln T}{d \ln M_{\rm{c}}} = b_{1} + b_{2} \frac{d \ln R_{\rm{c}} }{d \ln M_{\rm{c}}}. 
\end{align}
The terms $d \ln R_{\rm{c}}/d \ln M_{\rm{c}}$ reflect the dependency on the core's mass to its radius, and can be evaluated once we have a relation between the two. For the sake of intuition and simplicity, we can adopt the mass-radius relation under a polytropic approximation for degenerate matter based with index $n = 3$. This yields $R_{\rm{c}} \sim M_{\rm{c}}^{-1/3}$. \citep[See e.g.][]{chandrasekhar67}
 To make things more concrete, we assume Kramer's opacity and CNO energy generation as,
\begin{align}
\kappa \propto \rho T^{-2/7}, \; \; \epsilon \propto \rho T^{13}.
\end{align}
Under these assumptions, a homology analysis leads to $a_{1} = 7, a_{2} =-16/3$, while the coefficients that enter the scaling of temperature are independent of opacity and energy generation, and equal to $b_{1} = 1, b_{2} = -1$ (see e.g \citet{kippenhahn90}). Therefore, we have that,
\begin{align}
& T \sim  M_{\rm{c}} \cdot R_{\rm{c}}^{-1} \sim M_{\rm{c}}^{4/3}, \; \; \; L \sim  M_{\rm{c}}^{39/4} \cdot R_{\rm{c}}^{-31/4} \sim M_{\rm{c}}^{37/3}.
\end{align}
Therefore, the temperature in the shell and total luminosity of the star increase strongly with core mass. Considering that the growth of the core's mass with time is proportional to the luminosity provided by the shell, we can write  
\begin{equation}
M_{\rm{c}}  \frac{d \ln M_{\rm{c}}}{dt}  \propto L[M_{\rm{c}} ; \ldots],  \label{eq:dMdt}
\end{equation}
i.e, the rate of mass accumulation in the core will increase more and more with the accompanying increase in the luminosity. In the above relation, the ellipses in the functional argument of luminosity denote the implicit dependence of luminosity on input physics (opacity, nuclear rates, etc.). These fine details will determine {\it both} the rate of growth of $M_{\rm{c}}$, as well as the temperature $T$ at the surface of the degenerate core, and in turn, {\it the point along the Herzprung-Russell diagram} at which helium ignition in the degenerate core ($T \simeq 10^{8}$K) will be reached. The latter, signals the onset of the TRGB. 

We now ask the question of how the above estimates will vary under some uncertainty in the opacity and/or nuclear energy generation. To answer this, we will stick to our previous simplified picture, and model such uncertainties as a small perturbation in the scaling of relevant quantities. So, assuming the perturbed opacity and energy generation\footnote{For simplicity, we perturb only the exponent of temperature as it is the driving term in both expressions.} as $\epsilon \propto \rho T^{13 + \epsilon_1}, \; \; \kappa \propto \rho T^{-2/7 + \epsilon_2}$, where $\epsilon_1, \epsilon_2 \ll 1$, the luminosity equation becomes
\begin{equation}
\frac{d \ln L}{d \ln M_{\rm{c}}} \simeq \frac{1}{3} (37 + 2 \epsilon_1 - 2 \epsilon_2). \label{eq:dLdM_2}
\end{equation}
As can be seen, a small increase in the energy generation ($\epsilon_1 > 0$) will act as to increase the luminosity, whereas an increase in opacity ($\epsilon_2 > 0$) will have the opposite effect. Clearly, this is an oversimplified picture which cannot capture accurately the actual physics. Moreover, it cannot allow for an estimate of the effect of the conductive opacity in the degenerate core. However, on qualitative grounds, the conductive opacity will affect the strength of the temperature gradient in the core, and thus its temperature $T_{\rm{c}}$. 
At the same time, loss of energy in the form of neutrinos would add a term acting in a similar direction with $\epsilon_2$ in (\ref{eq:dLdM_2}) towards a reduction of the luminosity.

\section{Description of input physics and numerical procedure} \label{sec:Assumptions}
The models are constructed using the \code{} stellar evolutionary code \citet{tognelli22}, where a full exposition of the code's features can be found. Here, we limit ourselves to the description of the the main input physics/quantities and their reference values. The effect of variation of such quantities on the TRGB luminosity are discussed in the next section.

\begin{itemize}
\item Equation of State (EOS). The EOS describes the relation between temperature, density and pressure and closes the system of stellar evolution equations. Throughout our analysis, we employ the most recent version of the OPAL EOS \citep{rogers02} released in 2006 on the OPAL website\footnote{\url{https://opalopacity.llnl.gov/EOS_2005/}}.\\

\item Radiative opacity (\krad). Radiative opacity affects the way energy is transferred through radiation in the relevant stellar regions. Close to the TRGB, it affects the energy transport at the hydrogen burning shell around the helium core, and thus, the temperature at which helium ignites. We use Opacity Project tables \citep[OP,][]{badnell05} for a mixture of hydrogen, helium and metals and the OPAL 2005 Type-2 radiative opacity tables\footnote{\url{https://opalopacity.llnl.gov/new.html}.} for pure elements with enhanced Carbon and Oxygen \citep{iglesias96}.\\

\item Conductive opacity (\kcon). It affects the transfer of energy in the degenerate helium core, and therefore, it becomes important close to the TRGB. We adopted the \citet{potekhin99} tables available on online\footnote{\url{http://www.ioffe.ru/astro/conduct/}}.\\

\item Metal abundances. We assume solar scaled metal abundances\footnote{The abundances of metals in the star have been obtained by scaling the solar abundances: thus for any metal $i$ its numerical abundance in the star is evaluated as $N_{\rm{i}}=N_{\rm{i},\sun} Z/Z_{\sun}$.} form \citet{caffau11} completed for the missing elements by \citet{lodders09} abundances. \\

\item Nuclear reaction rates. From the main sequence up to the TRGB, the relevant reaction rates are p+p (proton-proton), p+d (proton-deuterium), $^3\rm{He}+^4\rm{He}$, $^3\rm{He}+^3\rm{He}$, $^7\rm{Be}+e^{-}$,  $^7\rm{Be}+\rm{p}$, CNO, $^{14}\rm{N}$+p, 3$\alpha$. Each reaction network becomes important at a different regime.  The pp-chain affects the hydrogen burning in MS, the CNO both the MS and the RGB shell hydrogen burning, while the $3\alpha$ reaction is the relevant one for the helium burning in the degenerate core during the TRGB phase. As we will discuss later, the stage at which helium ignites mainly depends on the efficiency of CNO and $3\alpha$ reactions (see Sect.~\ref{sec:rates}).\\

\item Electron screening. Electrons in the nuclear plasma tend to reduce the Coulomb barrier between two interacting charge particles, thus enhancing the nuclear burning reaction rates by a factor $f$. In the code, the screening is automatically evaluated using the formalism given in \citet{graboske73} and \citet{dewitt73}. Electron screening is important both during the main sequence and the TRGB phase to properly evaluate the rate of nuclear burning and so the grow (in mass) of the helium core. \\

\item High temperature neutrinos production. As the star approaches the Tip, in the center of the core the production of thermal neutrinos becomes efficient. Such particles weakly interact with the surrounding matter carrying away part of the energy, thus cooling the inner regions of the helium core. To treat the neutrinos productions we used the formalism given in \citet{itoh89} (photo-neutrinos and pair-production), \citet{haft94} (plasma-neutrinos) and \citet{dicus76} (bremsstrahlung neutrinos).\\

\item Outer boundary conditions. The pressure and temperature at the bottom of the atmosphere are consistently evaluated by integrating a hydrostatic atmosphere using a $T=T(\tau,T_\mathrm{eff})$ relation according to the prescription of \citet{vernazza81}.\\

\item Element diffusion. The diffusion of all elements is taken into account by solving for the diffusion velocities according to the formalism of \citet{thoul94}. Diffusion of elements throughout the star's evolution become important in view of their effect on quantities such as the radiative opacities and the mean molecular weight.\\

\item Mass loss. Throughout the evolution from the Main Sequence (MS) phase to the TRGB, the star is expected to lose a fraction of its mass, impacting its effective temperature and radius. The mass loss effect depends on the initial mass and metallicity of the star. To model this effect, we adopt the formulation of \citet{reimers75} which depends on a single parameter $\eta$, controlling the significance of the effect. For our reference models we use the value $\eta = 0.3$, which has been shown to be a typical value for the initial mass range we consider in this work \citep[see e.g.][]{hidalgo18}.\\

\item Core overshooting. During the MS, stars may develop a convective core depending on their mass and chemical composition. In these cases, the presence of overshooting leads to an increase of the hydrogen burning convective core and in turn to a more massive helium core at the end of MS (hydrogen exhaustion). However, as most of the models in our grid have a radiative core in MS (i.e. for $M\la 1.2$~\msun), we choose to set core overshooting to zero throughout our set of reference models. This assumption does not hinder generality of results, because even for the mass range where core overshooting could be present (1.2-1.4~\msun), MS convective cores are relatively small and the inclusion of a certain level of overshooting has only a small effect on the helium core at the end of the MS (see Sec.~\ref{sec:over}).  \\

\item Super-adiabatic convection. We implement the mixing-length theory \citep[MLT,][]{bohm58} to evaluate the heat transfer and temperature gradient in convective regions. Within MLT, the mixing length scale is proportional to the pressure scale height, with proportionality constant the so--called mixing-length parameter \ml. For the latter, we use the rounded value of our solar calibrated mixing length parameter, \ml=2.00. \\
\end{itemize}

\begin{figure}
\centering
\includegraphics[width=0.98\columnwidth]{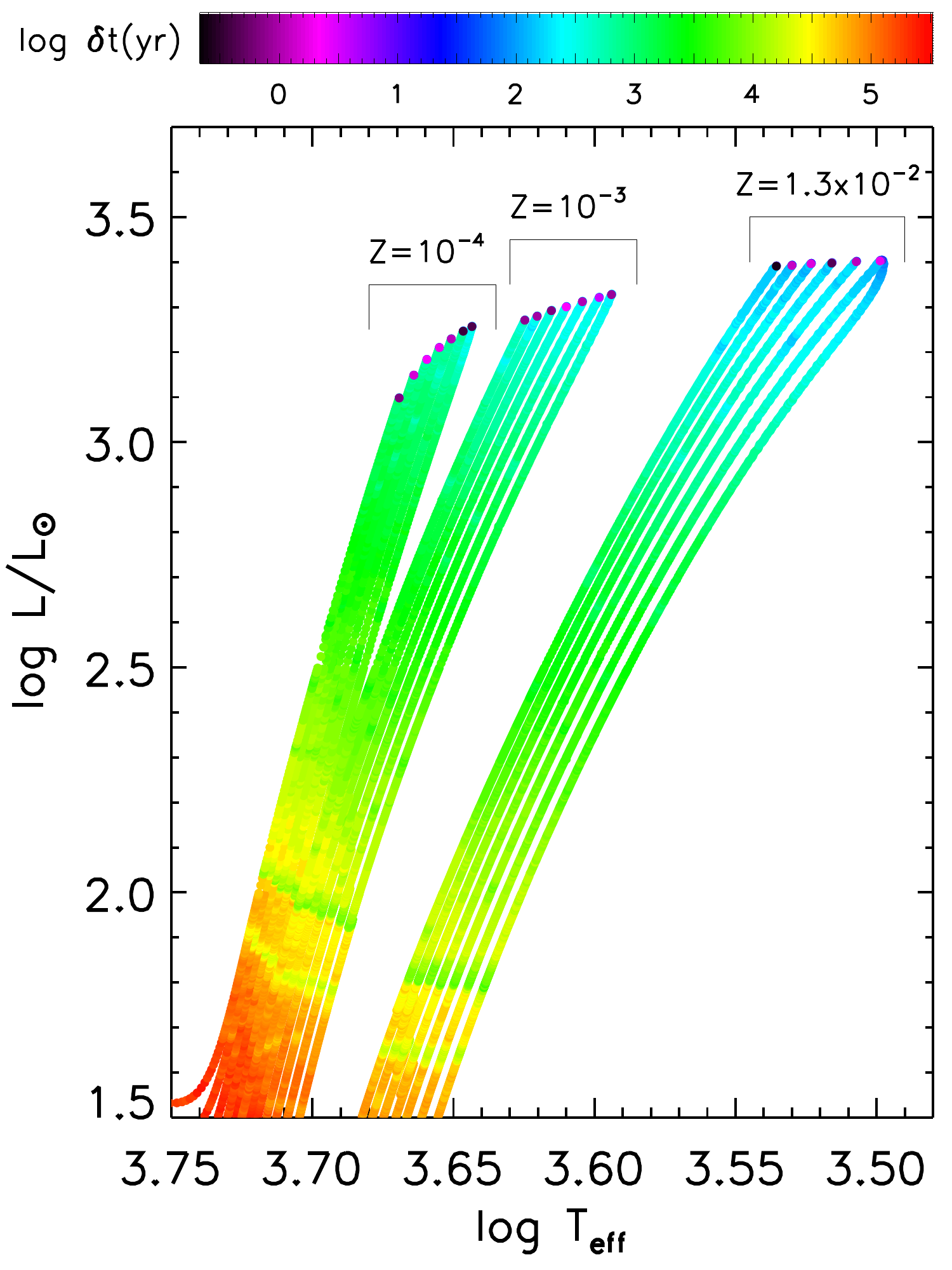}
\caption{Reference evolution along the red giant branch (RGB) on the HR diagram, for three metallicities $Z=0.0001, 0.0010, 0.0130$, and for masses in the range $[0.8,1.4]$~\msun. The color indicates the value of the models' time step ($\log \delta t$(yr)). For each value of stellar mass, we plot only about 4\% of the models in order to make the figure more accessible.
}
\label{fig:time_step}
\end{figure}
\begin{figure}
\centering
\includegraphics[width=0.98\columnwidth]{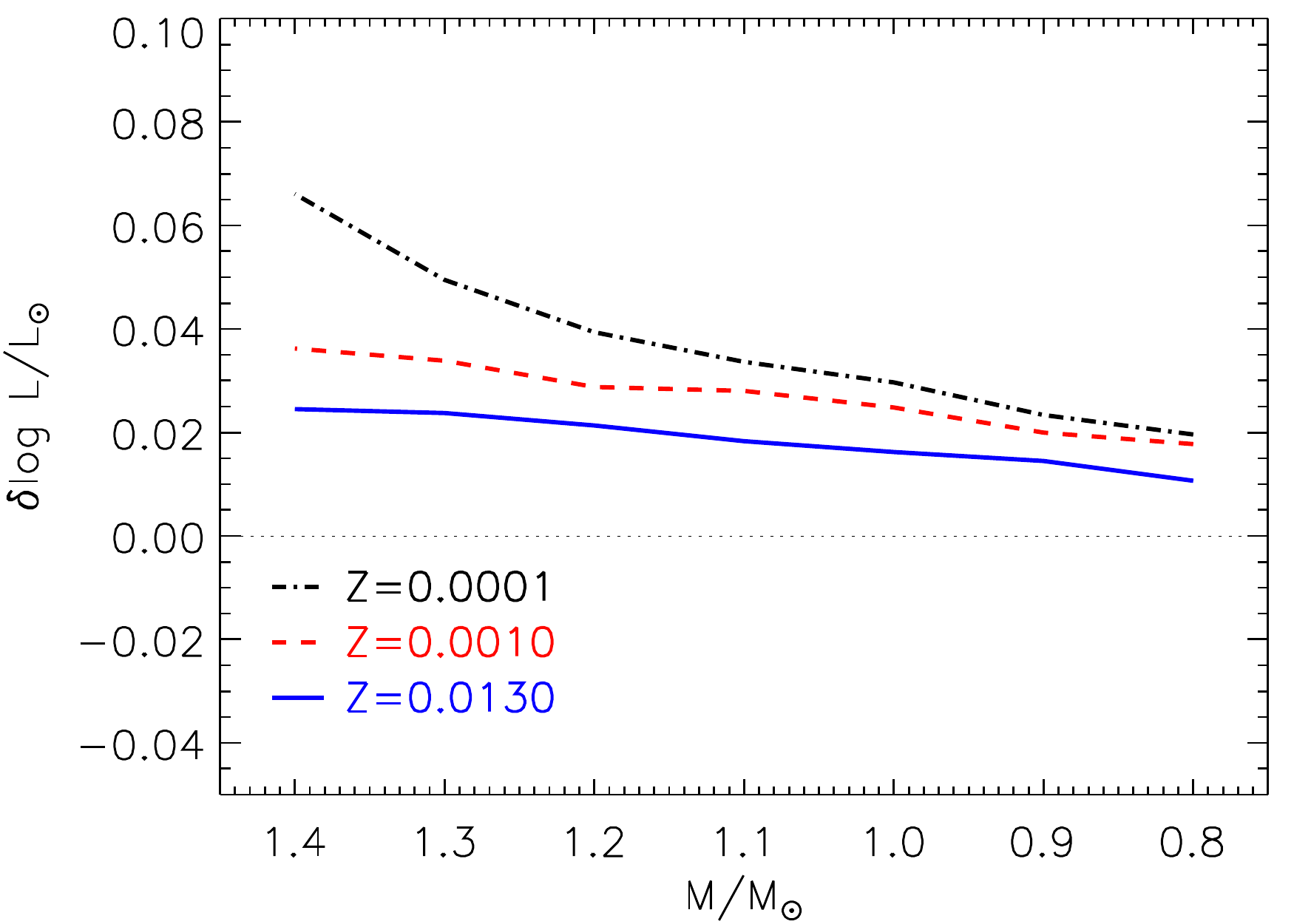}
\caption{Variation of the Tip luminosity induced by the adopted temporal resolution in RGB (time step) in the sense (not-optimised temporal resolution) - (optimised temporal resolution). The models are computed for the mass range [0.8,1.4]~\msun{} and for the three labelled overall metallicities.}
\label{fig:time_step2}
\end{figure}

\subsection{Computational accuracy: TRGB dependence on adopted time step}
\label{sec:timestep}
As a first step before the computation of the grid of models, we evaluate the stability of the results under a variation of the temporal resolution during the evolution through the RGB phase. Since the luminosity of the TRGB depends on the mass of the helium core at the flash, which in turn is affected by the rate at which it grows during the RGB phase, the choice of temporal resolution is important. In particular, if it is not high enough, it is possible that the luminosity of the Tip is affected on some level. This issue has been partially discussed in \citet{serenelli17}, where it was claimed that the adoption of a low enough temporal resolution during the RGB (i.e. large value of the time step) introduces a systematic effect on the predicted Tip luminosity, leading to an error of about 0.02~dex. In our simulation, in order to properly follow the evolution of the helium core in the RGB phase we implemented in the adaptive time step subroutine a control to limit the relative variation of the mass of helium core to less than $10^{-6}$ between two consecutive models. After trial and error, we found that this constrain (hereafter, ``optimised" time step) avoids uncontrolled fluctuations of the TRGB luminosity due to a variation in the adopted time step, i.e. when perturbing the parameters. With this additional constraint on the time step, the RGB evolution and in particular the growth in mass of the helium core, is carefully followed. For the sake of concreteness, the adopted time step ($\delta \log t$) for RGB stars linearly decreases from about $3 \times 10^5$~yr at the bottom of RGB ($\log L$/\lsun$\sim 1.5$) to about  $10^4$~yr at $\log L$/\lsun$\sim 2$, $300$~yr at $\log L$/\lsun$\sim 3.0$ and it continues to decrease quickly close to the helium flash, passing from $100$~yr to $10^{-1}$~yr in less than $\delta \log L$/\lsun$\sim 0.01$~dex close to the Tip. This way, each evolutionary track contains 8000-30\,000 RGB models, with the number of models increasing (decreasing) with the mass (metallicity). 

Figure~\ref{fig:time_step} shows the evolution of the time step along the RGB for stellar models with a mass $M\in[0.8,1-4]$~\msun{} for three values of the total metallicity ($Z=$0.0001, 0.0010 and 0.0130). We analysed the impact on the Tip luminosity of the adoption of a not-optimized time step, finding that it can result in a systematic overestimation of the Tip luminosity as large as 0.01-0.06~dex. In this regard, Figure~\ref{fig:time_step2} shows the difference in the Tip luminosity between models computed assuming the not-optimised adaptive time step, and the optimised one for the helium core evolution in RGB. In the first case, the time step in RGB (for $\log L$/\lsun$\ga 2$) is on average about 2.5-3.5 times larger than that in the second one. The largest effect on the Tip occurs in metal poor models, where the time step is systematically larger than in metal rich models. It is clear that the adoption of an optimised time step is thus mandatory to have a reliable and numerically stable prediction for the TRGB luminosity. 

\begin{figure*}
\centering
\includegraphics[width=0.32\linewidth]{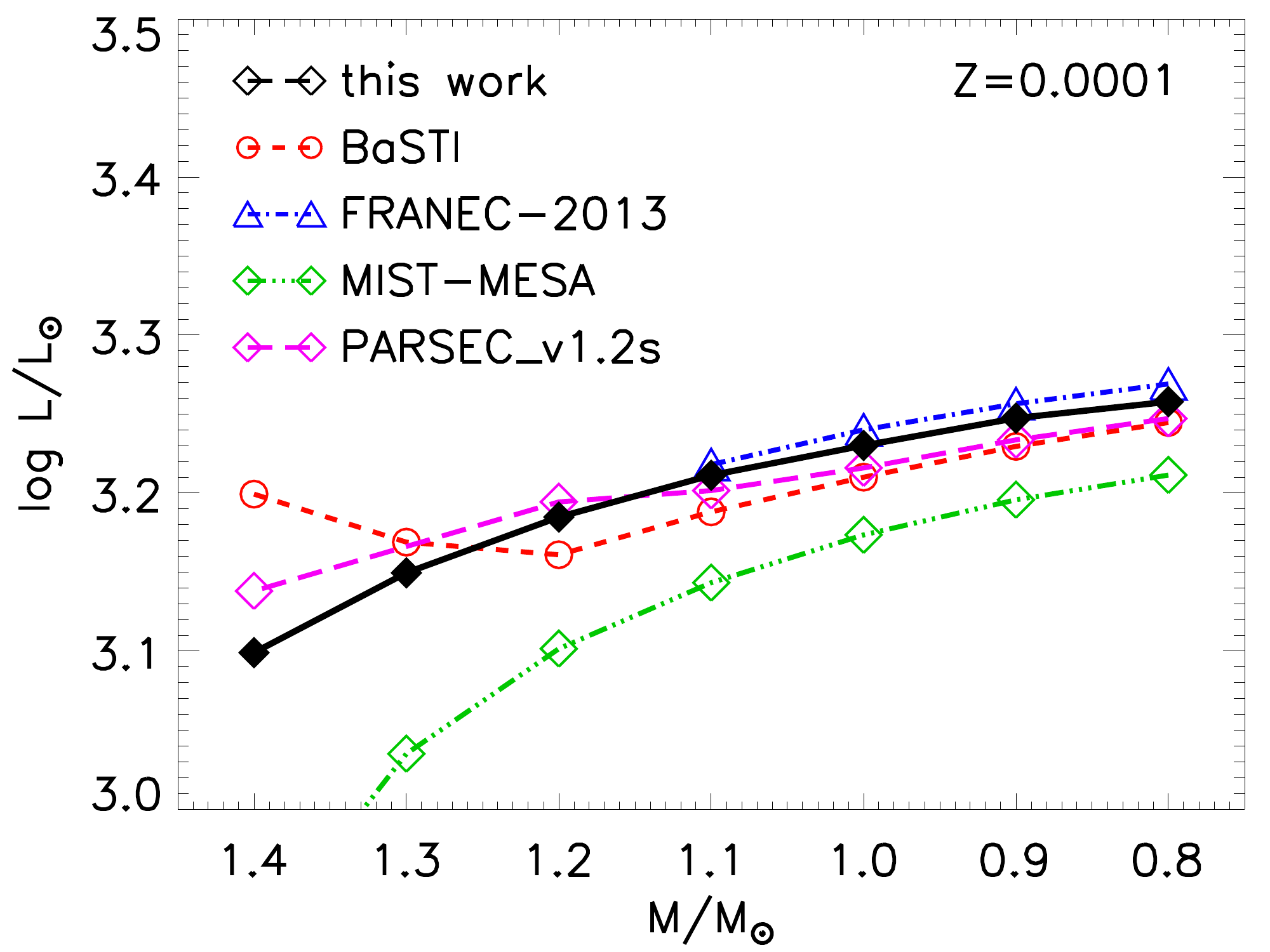}
\includegraphics[width=0.32\linewidth]{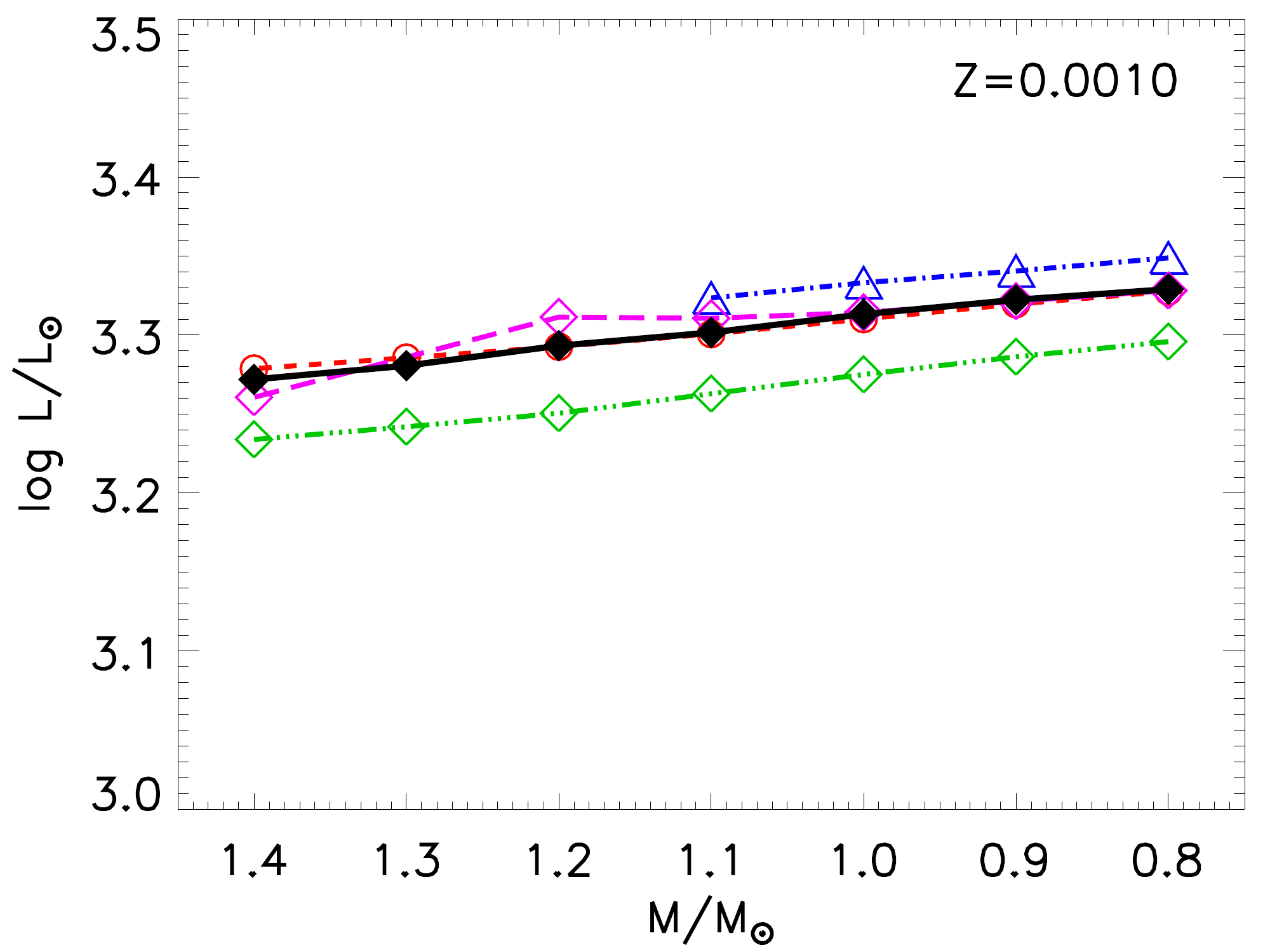}
\includegraphics[width=0.32\linewidth]{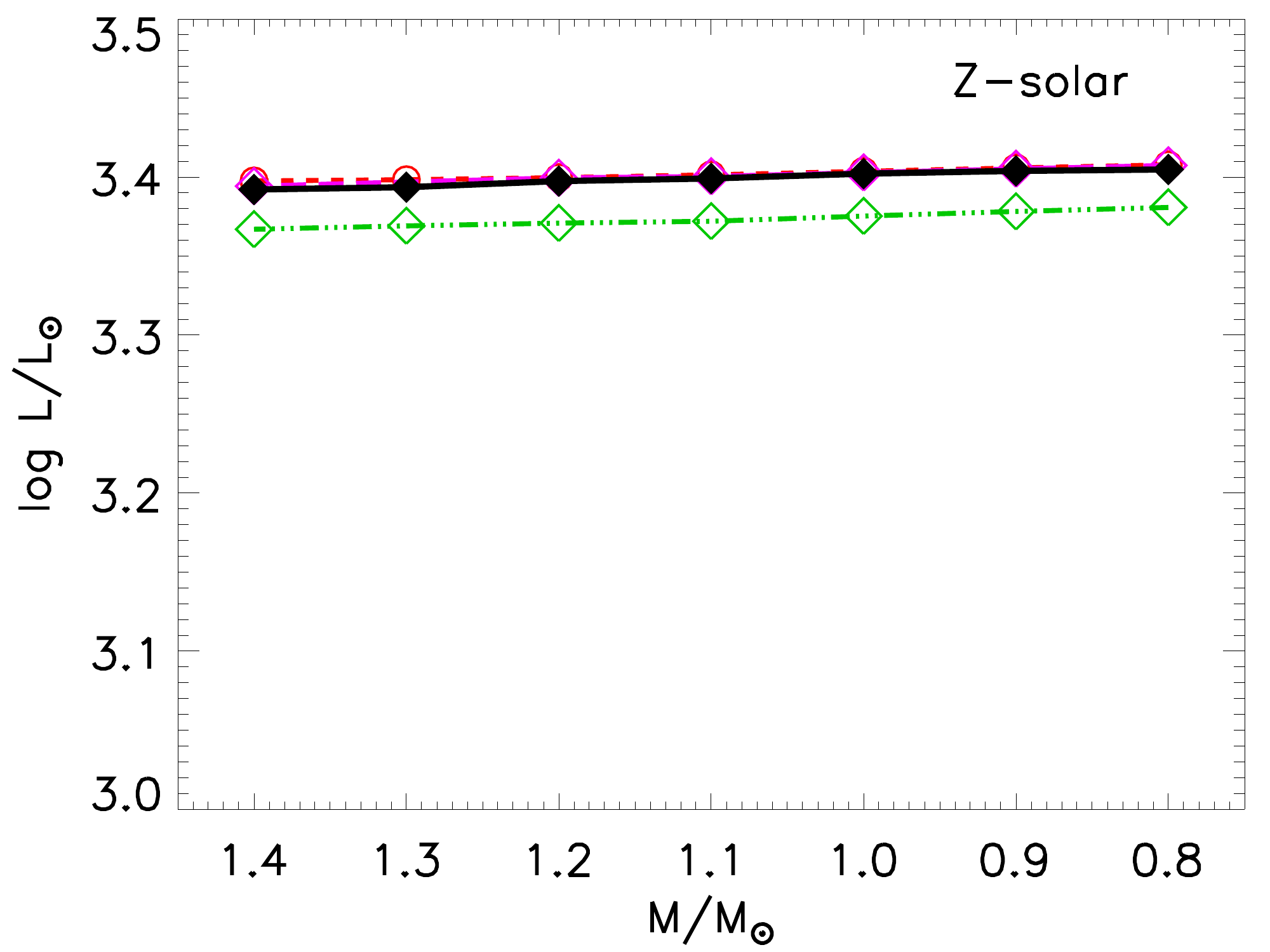}
\caption{Comparison between the TRGB luminosity predicted by the quoted stellar codes at three metallicities: in the case of solar metallicity FRANEC-2013 models are not available while the other models adopt Z=0.0130 but MIST-MESA which uses Z=0.0140.}
\label{fig:comp}
\end{figure*}

Figure~\ref{fig:comp} shows the Tip luminosity predicted by our reference set of models compared to that predicted by other models extracted from the following stellar databases: BaSTI \citep{hidalgo18}, FRANEC \citep{deglinnocenti08,dellomodarme12,valle13}, MIST-MESA \citep{choi16,dotter16}, PARSEC\_v1.2s \citep{bressan12,chen2014}. The overall agreement between the selected models is good for all the selected metallicity with the exception of MIST-MESA models that systematically underestimate the Tip luminosity of about 0.03~dex at solar metallicity, increasing to about 0.04~dex for Z=0.0010 and to 0.05-0.1~dex for Z=0.0001. The other models show an excellent agreement at solar metallicity, while at lower metallicities  differences increase to about 0.02~dex at sub-solar metallicities for most of the selected stellar mass values. It is interesting to notice the behaviour of the BaSTI models at Z=0.0001 for $M\ga 1.2$~\msun, in the regime of low-electronic degeneracy in the helium core -- there, the difference with the other models increases to about 0.1~dex in the worst case. We cannot find a firm explanation for this behaviour.

\section{Effect on the TRGB luminosity under individual perturbations of each quantity} \label{sec:uncertainties}
As a preliminary analysis of the impact of input physics on the total error on the TRGB luminosity, we start by varying {\it one input parameter at a time}. Our goals in this section are the following:
\begin{itemize}
\item To understand both the quantitative and qualitative effect of each individual input physics parameter on the TRGB, in a relatively simple and intuitive manner, before our full grid analysis.
\\

\item  To analyse the linear response of the TRGB luminosity, under individual variations of each input physics parameter. As we will argue, the TRGB responds linearly as long as the evolved models also respond linearly within the input physics parameter range we are interested in. Later on, we will re-visit this result in view of a Monte-Carlo analysis. 
\end{itemize}

We recall here that, the luminosity of TRGB is strongly correlated to the mass of the helium core in the star. Therefore, any quantity that affects the growth in mass of the helium core or the onset of He burning at the flash directly affects the Tip luminosity. The quantities that we took into account in this work are listed in Table~\ref{tab:pert} along with the perturbation that we applied while the variation of the Tip luminosity caused by such perturbed quantities is shown in Table~\ref{tab:pert_eff}.  Notice that, the Table~\ref{tab:pert_eff} has been divided in three parts: the first part lists the parameters that produce an effect on the Tip luminosity of at least $10^{-3}$~dex, the second part the parameters that produce a smaller effect, and the third part the parameters for which we cannot firmly associate an error estimate (i.e. mass loss parameter, core overshooting, electron screening, \ml). We also found that in the cases where the variation of the Tip luminosity is smaller than $10^{-3}$~dex, numerical fluctuations due to the adopted time step (see Sec.~\ref{sec:timestep}), could introduce spurious and uncontrolled effects in the Tip model. For this reason, these quantities have been separated from the others.

\begin{table*}
\centering
\caption{List of the quantities analysed in this work with the related uncertainty.}
\begin{tabular}{lrr}
\hline
quantity ($G_j$) & perturbation ($\delta_{G_j}$) & references\\
\hline
\hline
Radiative opacity (\krad) &  $\pm$5\% & VD13, T18, S20\\ 
Conductive opacity (\kcon) & $\pm$5\% & VD13, S20\\
$^{14}$N(p,$\gamma)^{15}$O & $\pm$10\% & LUNA05, VC13, VD13, S20\\
$\alpha(\alpha\alpha,\gamma)^{12}$C & $\pm$20\% & NACRE99, VC13, VD13, S20\\
$\rm{p}(p,e^+\nu)d$ &  $\pm$1\% & AD11, MSV13 \\
$^3\rm{He}(^3He,2p)\alpha$ &  $\pm$4\% & AD11\\
$^3\rm{He}(\alpha,\gamma) ^7$Be & $\pm$7\% & AD11\\
$^7\rm{Be}(e^-,\nu)^7$Li & $\pm$ 2\% & AD11 \\
$^7\rm{Be}(p,\alpha)\alpha$ & $\pm$10\% & AD11 \\
Electron screening ($f_\mathrm{sc}(3\alpha)$) & weak & (see text) \\
Electron screening ($f_\mathrm{sc}(^{14}\mathrm{N})$) & weak & (see text) \\
Neutrino energy loss ($\nu$) & $\pm$5\% & VC13, VD13, S20\\
Outer boundary conditions (BCs) & $\pm$5\% & VC13, S20\\
Diffusion coefficients ($v_{\rm{dif.}}$) & $\pm$15\% & TBL94, VD13\\
Core overshooting ($\beta_\mathrm{ov}$) & 0, 0.15 & (see text) \\
Mixing length (\ml) & $\pm0.2$, $\pm 0.4$ & (see text) \\
Mass loss ($\eta_\mathrm{Reim.}$) & $0, 0.2, 0.3, 0.4$ & (see text) \\
Primordial helium abundance ($\delta Y_p$) & $\pm 0.0015$ & (see text) \\
helium enrichment, $\delta(\Delta Y/ \Delta Z)$ & $\pm 1$ & (see text) \\
\hline
\end{tabular}
\\
\footnotesize 
Refs.: AD11 \citep{adelberger11}; LUNA05 \citep{imbriani05}; MSV13 \citep{marcucci13}; NACRE99 \citep{angulo99}; VC13 \citep{viaux13}; VD13 \citep{valle13}; T18 \citep{tognelli18}; S20 \citep{straniero20}; TBL94 \citep{thoul94}
\normalsize
\label{tab:pert}
\end{table*}
\begin{table*}
\centering
\caption{TRGB luminosity variation ($\delta \log L/$\lsun) due to the perturbation of each quantity ($\delta_{G_j}/G_j$) related to input physics. The explicit description of each quantity is given in Table~\ref{tab:pert}. We divided the table in three parts: the first with those quantities that affect the TRGB log-luminosity at least by $10^{-3}$, the second with an effect smaller than $10^{-3}$ for all the selected values of $Z$, and a third part where we list the luminosity variation of input physics  for which we did not assign a value for the $\sigma$ error (see text). The flag ``ref." means that the parameter is not varied, but its reference value is used. The last column contains a flag to indicate if the quantity has been taken into account in the cumulative error evaluation (1) or not (0), that is, the case where all considered parameters are perturbed simultaneously when computing the impact on the Tip luminosity.
}
\begin{tabular}{lrrrrr}
\hline
$G_j$ & $\delta_{G_j}/G_j$ & \multicolumn{3}{c}{$\sigma_L \equiv \delta \log L/$\lsun} & flag\\
 & & $Z=0.0001$ & $Z=0.0010$ & $Z=0.0130$ \\
\hline	
\hline %
$\kappa_{\rm{rad.}}$ & $\pm  0.05$ & $\mp 0.009$ & $\mp 0.011$ & $\mp 0.014$ & 1\\
$\kappa_{\rm{con.}}$ & $\pm  0.05$ & $\mp 0.004$ & $\mp 0.004$ & $\mp 0.003$ & 1\\
$^{14}$N(p,$\gamma)^{15}$O & $\pm 0.10$ & $\pm 0.003$ & $\pm 0.003$ & $\pm 0.002$ & 1\\
$\alpha(\alpha\alpha,\gamma)^{12}$C & $\pm 0.20$ & $\mp 0.008$ & $\mp 0.006$ & $\mp 0.005$ & 1\\
$\nu$ & $\pm  0.05$ & $\pm 0.008$ & $\pm 0.006$ & $\pm 0.004$ & 1\\
$\mathrm{BCs}$ & $\pm  0.05$ & $\pm 0.001$ & $\pm 0.002$ & $\pm 0.001$ & 1\\
$Y$ & $\pm  1\sigma$ & $\mp 0.001$ & $\mp 0.001$ & $\mp 0.002$ & 1\\
\hline
$v_{\rm{dif.}}$ & $\pm 0.15$ & $\pm 0.0004$ & $\pm 0.0004$ & $\pm 0.0003$ & 0 \\
$\rm{p}(p,e^+\nu)d$ & $\pm  0.01$ & $<10^{-4}$ & $<10^{-4}$ & $<10^{-4}$ & 0\\
$^3\rm{He}(^3He,2p)\alpha$ & $\pm  0.04$ & $<10^{-4}$ & $<10^{-4}$ & $<10^{-4}$ & 0\\
$^3\rm{He}(\alpha,\gamma) ^7$Be & $\pm  0.07$ & $<10^{-4}$ & $<10^{-4}$ & $<10^{-4}$ & 0\\
$^7\rm{Be}(e^-,\nu)^7$Li & $\pm  0.02$ & $<10^{-4}$ & $<10^{-4}$ & $<10^{-4}$ & 0\\
$^7\rm{Be}(p,\alpha)\alpha$ & $\pm 0.10$ & $<10^{-4}$ & $<10^{-4}$ & $<10^{-4}$ & 0\\
\hline
\ml & $\pm0.20$ & $\mp 0.0003$ & $\mp 0.0006$ & $\mp 0.0007$ & 0\\
$\beta_\mathrm{ov}$~$^{(a)}$ & ref./0.15 & 0.004-0.037 & 0.004-0.013 & 0.001-0.002 & 0\\
$\eta_\mathrm{Reim.}$ & ref./0 & $< 0.001$ & $< 0.001$ & $ < 0.001$ & 0\\
$f_\mathrm{sc}(3\alpha)$ & ref./weak &  $-0.031$ & $-0.027$ &  $-0.021$ & 0\\
$f_\mathrm{sc}(^{14}\mathrm{N})$ & ref./weak & $<0.001$ & $<0.001$ & $<0.001$  & 0\\
\hline
\footnotesize $^{(a)}$ $\beta_\mathrm{ov} =0.15$ only for $M\ge 1.2$~\msun. \normalsize
\end{tabular}
\label{tab:pert_eff}
\end{table*}

\subsection{Radiative and conductive opacities}
A firm evaluation of the uncertainties on opacity coefficients (\krad) is not available, although in the literature a conservative uncertainty of $\pm $5\% is usually adopted \citep[see e.g.][]{valle13,tognelli18,straniero20}. Such a value is derived as the relative difference between different opacity tables. We estimated the uncertainty on the radiative opacity \krad{}, comparing our reference opacity table (i.e., the OP) against another popular opacity tabulation, known as OPAL \citep{iglesias96} updated in 2005. We verified that in most of the temperature-density domain typical of the evolution of RGB stars, the maximum relative differences between radiative opacity is of the order of $\pm 5$\% for $5.6 \la\log T\la 7.8$, while it can reach larger values (up to 10-15\%) in the convective envelope ($\log T \la 5.6$) or at the stellar center, for $\log T\ga 7.8$. However, in the center of RGB stars where pressure support is provided by the degenerate state of matter, the energy transport occurs mainly by electron conduction, and consequently the uncertainty on radiative opacity is negligible. On the other hand, the effect of varying the radiative opacity in the envelope affects only the $T_\mathrm{eff}$ of the star, since it controls the transfer of energy from the hydrogen burning shell to the surface. Thus, the only region where the radiative opacity can have a significant effect on the Tip luminosity is close to the boundary between the helium core and the convective envelope, where there is a competition between the electron conduction and radiative energy transport. In this region the efficient hydrogen burning shell causes the helium core's mass to grow. A variation of the energy transport in this part of the star can potentially affect the temperature of the helium core and consequently accelerate/postpone the occurrence of the helium flash, thus resulting in a variation of the Tip luminosity.

We found that the radiative opacity is one of the main uncertainty sources of the Tip luminosity producing a luminosity variation between 0.010-0.015~dex, as shown in Table~\ref{tab:pert_eff}. The importance of radiative opacity in determining the Tip luminosity has been debated in the literature. \citet{viaux13} found that the radiative opacity does not contribute significantly to the uncertainty of the Tip, a result that has been confirmed by \citet{straniero20}. On the other hand, \citet{valle13} found that a variation of the radiative opacity of $\pm 5$\% affects the Tip luminosity at a level of $10^{-2}~$dex, the same effect we found in our analysis. It is not possible to understand such differences without having information on how the radiative opacity has been perturbed in a given stellar model. In our work, we answer this question by quantifying the effect of radiative opacity on Tip luminosity through a variation of \krad{} in different stellar regions. 

As expected, a variation of the  radiative opacity inside the helium core does not produce any appreciable effect, because the dominant type of energy transport is via electron conduction. Similarly, we verified that a variation of \krad{} only for $T\le 10^6$~K, i.e. in the envelope, affects only the star's effective temperature, leaving no effect on the Tip luminosity. On the other hand, we found that varying the radiative opacity only in the temperature region where $2\times 10^7 \le T\le 7\times 10^7$ while leaving it unchanged in the other parts of the star, produces the entire Tip luminosity variation that we found. Therefore, and as anticipated, we found that the variation of the radiative opacity at the top of the helium core affects the temporal evolution of its temperature. In particular, the larger the opacity, the hotter the core ends up being at a given time, and consequently the smaller the mass of the helium core (and total luminosity $L$) at the flash. Therefore, increasing (decreasing) the radiative opacity decreases (increases) the LTRGB (see also Sec. \ref{sec:theory}). We emphasize that, our result is in very good agreement with that found \citet{valle13}. 

Concerning conductive opacities (\kcon), the uncertainty on opacity calculation has been estimated to be about 5-10\%. \citet{viaux13} adopted an uncertainty of 10\% while \citet{valle13} suggested that an uncertainty of 10\% could be expected in the outer part of the helium core, while in most of the core one can expect an uncertainty of about 5\%. An uncertainty of $\pm 5$\%, has also been adopted recently by \citet{straniero20}. Following \citet{valle13} and \citet{straniero20} estimations, we assumed a conservative uncertainty of 5\% in all the T-$\rho$ regime. The effect on the Tip luminosity are summarised in Table~\ref{tab:pert_eff}. As it turns out, the conductive opacity has an impact on the luminosity of about 3-4$\times 10^{-3}$~dex, hence much smaller than that of radiative opacity. The underlying reason behind this weak dependence is that, a variation of the conductive opacity predominantly affects the way the energy is re-distributed inside the helium core, whereas the energy transport outside the core, produced by the hydrogen-burning shell, is determined by the radiative opacity. Therefore, the effect of conductive opacities becomes secondary in this context. 

Before ending this section, it is worth mentioning that recently \citet{cassisi21} found that the uncertainty on conductive opacity could be much larger than 5\%, mainly in the regime of weak level of electronic degeneration. In our case, their results could affect mainly the uncertainties estimated for metal poor stars (i.e. $Z \la 0.001$) with $M\le 1$\msun, as at larger metallicities models in the mass range 0.8-1.4~\msun{} tend to have a large level of degeneracy in the helium core. However, we warn that this result could potentially overturn the relative importance of radiative/conductive opacities in metal poor cases.

\subsection{Nuclear reaction rates and electron screening}
\label{sec:rates}
The systematics from nuclear reactions come mainly from the reaction rates of the CNO and $3\alpha$ process. We describe each of them separately below, before we discuss the issue of electron screening.

The growth in mass of the central helium core is affected by the efficiency of nuclear reactions in the hydrogen-burning shell, which processes hydrogen into helium predominantly via the CNO cycle. A more (less) efficient reaction rate will lead to a larger (smaller) core mass in a given time interval, and in turn, leading to a larger (smaller) Tip luminosity. The efficiency of the CNO cycle is mainly controlled by the reaction rate of the $^{14}$N(p,$\gamma$)$^{15}$O channel which is the bottleneck of the whole cycle. Thus, we considered the possible effect due to an uncertainty on the latter channel. In particular, we adopted the \citet{imbriani05} reaction rate who also estimated the uncertainty to be of about 10\%. Assuming his uncertainty we found that the Tip luminosity is weakly affected, with a luminosity variation of about 0.002-0.003~dex.

The TRGB luminosity is further affected by the $3\alpha$ reaction, $\alpha(\alpha\alpha,\gamma)^{12}$C, which is responsible for practically setting the value of the core mass at the helium-flash. In particular, a larger (smaller) value of such a reaction rate favours the ignition of helium earlier (later), hence leading to a smaller (larger) Tip luminosity. We adopted the reaction rate provided by NACRE, with an uncertainty of 20\%. The total effect on the Tip luminosity is about 0.005-0.007~dex in the selected metallicity range.

We also investigated how the Tip luminosity is affected by the hydrogen burning rate in MS. For most of the stars analysed in this work, the hydrogen burning in MS occurs mainly via pp-chain, which involves the following hydrogen burning channel: p+p, p+d, $^3\rm{He}+^4\rm{He}$, $^3\rm{He}+^3\rm{He}$, $^7\rm{Be}+e^{-}$ and  $^7\rm{Be}+\rm{p}$. The adopted reaction rates and the relative uncertainties are given in Table~\ref{tab:pert}. We verified that these burning channels have a very small impact on the Tip luminosity, producing a luminosity variation, at the Tip smaller than about $10^{-4}$, in good agreement to what found by \citet{viaux13}. 

Another aspect related to the nuclear burning that deserves to be discussed is the electron screening effect. Electrons are free to redistribute around charge nuclei, producing a screening effect that reduces the Coulomb barrier. Therefore, along with the measured uncertainty on the bare nucleus reaction rate, the actual rates in a star depend also on the efficiency of electron screening. The enhancement of the bare nucleus reaction rate due to screening is parametrised by the factor $f_\mathrm{sc}$. We considered the variation of the electron screening only for the two reactions that affect the TRGB, namely $^{14}$N(p,$\gamma$)$^{15}$O and and $\alpha(\alpha\alpha,\gamma)^{12}$C. 

In the models, the electron screening has been evaluated following the formalism presented in \cite{graboske73} and \citet{dewitt73}. In particular, for the $^{14}$N(p,$\gamma$)$^{15}$O reaction, we have checked that, in our simulations the value of the electron screening enhancement factor $f_\mathrm{sc}(^{14}\mathrm{N})$ varies between $\sim 1.2$ in the MS core, to $\sim 1.02$ in the RGB hydrogen burning shell (for $M\in[0.8,1.0]$~\msun). 

Having no firm uncertainties on the electron screening, we tried to estimate the effect of it on the predicted TRGB by changing the functional form of the screening function $f_\mathrm{sc}(^{14}\mathrm{N})$, setting it equal to the Salpeter weak-screening formula \citep{salpeter54}. The differences resulting from the adoption of these two different forms for the screening factor are relevant mainly during the MS phase, where the adoption of the Graboske \& De Witt or Salpeter formalism leads to a relative variation as large as $20\%$, with the Salpeter screening producing the larger value of $f_\mathrm{sc}(^{14}\mathrm{N})$. However, as the hydrogen burning proceeds off the stellar center, the difference between the two adopted formalisms reduces. 
As a consequence of this fact, the effect of $f_\mathrm{sc}(^{14}\mathrm{N})$ on the LTRGB is limited to the process of formation of the helium core during the MS, and as such, the total impact on LTRGB (i.e. $< 10^{-3}$) is small compared to the other systematics . 

Concerning electron screening associated with the $3\alpha$ reaction, it  becomes important only when the star ignites helium, hence, close to the Tip. We remind here that, the $3\alpha$ reaction involves two channels.  First, it requires the production of $^8$Be via the $\alpha+\alpha$ channel, and then, the production of $^{12}$C via the $^8$Be$+\alpha$ channel respectively. Therefore, the net electron screening is the product of the electron screening coefficients of the last two channels. The screening for this reaction progressively reduces from  $f_\mathrm{sc}(3\alpha)\approx 3$ to about 1.5-2 in the last part of the RGB when helium starts to ignite\footnote{To have a representative value of $f_\mathrm{sc}(3\alpha)$, we selected models in the RGB where the helium burning luminosity $L($He) is larger than $10^{-4}$ the stellar luminosity.}. Similarly to what done before, we evaluated the effect of adopting a Salpeter weak screening also for $3\alpha$ reaction: in this case, the value of $f_\mathrm{sc}(3\alpha)$ is overestimated in the Salpeter approximation of about 50\% at the Tip, while before the Tip $f_\mathrm{sc}(3\alpha)$ is overestimated by a factor larger than 2. Such a large variation of $f_\mathrm{sc}(3\alpha)$ causes a quite strong effect on the predicted Tip luminosity, with a luminosity change as large as 0.02-0.03~dex. Note that, since the use of the Salpeter weak regime overestimates the value of $f_\mathrm{sc}(3\alpha)$, in this case the ignition of helium occurs at a lower luminosity, i.e the TRGB luminosity reduces.

\subsection{Plasma neutrinos production}
As the helium core grows during the RGB phase, the large density and temperature in the inner regions of the helium core  make the production of thermal neutrinos efficient via 4 channels \citep[see e.g.][]{dicus76,itoh89,haft94}: 1) photo-neutrinos ($\gamma e^{-} \rightarrow e^{-}\nu\overline{\nu}$), 2) plasma-neutrinos ($\gamma \rightarrow \nu\overline{\nu}$), 3) pair-production ($e^{+}e^{-} \rightarrow \nu \overline{\nu}$), and 4) bremsstrahlung ($e^{-}(Ze^{-}) \rightarrow \nu\overline{\nu}$). Such particles weakly interact with the matter of the star, thus they practically travel throughout the star carrying away energy. The net effect of the production of thermal neutrinos is thus to cool the regions where they are produced, i.e. the center of the star. This mechanism is responsible for a temperature inversion in the helium core, where the maximum value of temperature is off center, in turn causing the helium flash to occur in shells. 

The production rate of neutrinos and the energy they carry away determine the cooling of the inner regions of the helium core, thus it directly affects the stage when helium is ignited, that is, the mass of the helium core and the Tip luminosity. A more (less) efficient production of neutrinos results in a colder (hotter) helium core. This directly translates into a more (less) massive helium core at the Tip and consequently in a brighter (fainter) Tip model.

To evaluate the effect of thermal neutrinos on the Tip, we adopted a commonly used uncertainty on thermal neutrinos energy loss of $\pm 5$\% \citep[see e.g.][]{valle13,viaux13,straniero20}. As shown in Table~\ref{tab:pert_eff}, such an uncertainty affects the Tip luminosity at a level of about $4-8\times 10^{-3}$~dex, the largest effect occurring in metal poor (hotter) models.

\subsection{Outer boundary conditions}
The equations that define the stellar model structure are integrated under two boundary conditions at the center and two at the surface of the star. Whereas the former one is fixed, relating to regularity of physical quantities at the stellar center (i.e. $R(M=0) = L(M=0) = 0$), the latter depends on the assumptions about the atmospheric layers. Usually the so--called $T-\tau$ relation is employed to relate temperature with opacity in the atmosphere, under which the atmospheric hydrostatic equation closes. Here, we study the influence of the outer boundary condition by varying the pressure and temperature at the bottom of the atmosphere with respect to that obtained with the reference $T-\tau$ relation \citep[][]{vernazza81}. We estimated the uncertainty by comparing the temperature and pressure at the bottom of the atmosphere obtained using the reference $T-\tau$ relation and those obtained using non-grey atmospheric models by \citet{allard11}, in the mass and chemical composition range used in this work. We found a variation of about $5\%$ on the temperature predicted at the bottom of the atmosphere, which we adopted as the $1\sigma$ uncertainty due to the BCs. As shown in Table~\ref{tab:pert_eff}, such an uncertainty changes the Tip luminosity by about $10^{-3}$~dex in the whole metallicity range.

\subsection{Diffusion velocities}
Microscopic diffusion has been taken into account during the whole evolution of models. The direct effect of diffusion on a model is to change the abundance profile along the stellar structure, with helium and metals that sink towards the center and hydrogen that moves upwards. However, diffusion plays a role only in radiative regions of stellar models and mainly during the MS phase. Indeed, pre-MS models are almost fully convective for most of their evolution and in conjunction with the relatively small time scale characteristic of this evolutionary stage, diffusion is not efficient enough to produce appreciable effects. A similar situation occurs in RGB, when the star has a thick convective envelope. On the other hand, during the MS the star develops thick radiative regions and the evolution (nuclear) time scale is large enough (in the selected mass range) to make diffusion efficient. From this qualitative description it is clear that, diffusion does not directly affect the Tip model, and its effect is to partially modify the helium core mass left at the end of the MS, without an effect on its growth during the RGB. 

To evaluate the impact of diffusion on the Tip, we adopted an uncertainty on diffusion velocities of $\pm 15$\%, as suggested by \citet{thoul94}. The variation of Tip luminosity is relatively small being at maximum of about 3-$5\times 10^{-4}$~dex, as shown in Table~\ref{tab:pert_eff}. Thus, diffusion is not one of the main uncertainty sources for the Tip luminosity.

\subsection{Mixing length parameter}
The super adiabatic gradient is evaluated using the MLT formalism, and as a reference case we have adopted \ml=2.0, which is an approximation of our solar calibrated value.  Concerning its uncertainty, being a free parameter, \ml{} does not have a formal error estimate. However, to assign a plausible variability range to it, we performed the following  test. We re-calibrated the value of \ml{} on the Sun by simultaneously perturbing \krad, $v_\mathrm{dif}$ and BCs, which are the quantities among those analysed in this paper, that mainly affect the solar radius, hence the calibration of \ml. We adopted the same perturbation discussed above, and we found a maximum variation of \ml{} of about $\pm0.2$, which we adopted as a conservative variability range around the reference solar calibrated one. The effect of a variation of \ml{} on the Tip luminosity is shown in Table~\ref{tab:pert_eff}, and in all the cases it is smaller than about $10^{-3}$~dex.
\subsection{Overshooting}
\label{sec:over}
The reference set of models, i.e without variation of the input physics parameters, has been computed without core overshooting. However, a certain level of overshooting in the core might be expected in those stars that develop during the MS phase a convective core. The value of the initial stellar mass for which core overshooting is expected to occur, and the amount of overshooting form two free model parameters. Concerning the transition mass, typically a value of 1.1-1.2~\msun{} is adopted, which corresponds to the mass range of stars which develop a convective core during the MS phase (for solar metallicity stars). The most suitable value for the amount of overshooting is still, however, debated. Core overshooting, which occurs predominantly during the MS phase, ends up increasing the accumulated helium mass in the core of the star, and hence, leading to a larger core mass close to the TRGB. We also recall here that, overshooting can be treated either as a diffusive process, or as a step increase of the extension of the core ($\ell_\mathrm{ov}$) by an amount proportional to the local pressure scale height, i.e. $\ell _\mathrm{ov} = \beta_\mathrm{ov} H_P$. In addition, it is not clear whether a constant value of overshooting parameter should be adopted or if a value that linearly increases with the total stellar mass should be used \citep[see e.g ][and references therein]{claret16}. Here, being interested in estimating the impact of the overshooting on the LTRGB, we adopted a simple step overshooting scheme, and we checked the impact on the Tip luminosity of adopting a constant value  $\beta_\mathrm{ov}=0.15$, for models with $M\ge 1.2$~\msun. The effect of the inclusion of overshooting depends on the stellar mass and chemical composition of the star. The mass of the helium core left at the end of the MS phase is larger in models with overshooting, and the impact increases with increasing the initial stellar mass. In Table~\ref{tab:pert_eff} we list the variation of the TRGB luminosity passing from $\beta_\mathrm{ov}=0$ (reference) to $\beta_\mathrm{ov} = 0.15$. The range reflects what stated before, that is, the effect of overshooting increases with increasing the stellar mass, producing a maximum luminosity variation from $\sim 4 \times 10^{-3}$ for $M=1.2$~\msun{} to values of about $3.7\times 10^{-2}$~dex for $M=1.4$~\msun. The effect on the luminosity is more important for metal poor models ($Z=0.0001$), because at a fixed mass the dimension of the hydrogen-burning convective core increases with decreasing metallicity.
\subsection{Mass loss}
Stars evolving along the RGB are expected to lose matter from their surface due to the coupling of radiation with cold and opaque matter as well as to the relatively low gravity. The effect is expected to increase with the luminosity of the RGB model. The mass loss is a very slow process, and consequently, it becomes important only in low-mass stars, which are those stars that spend enough time in RGB to produce an appreciable variation of the total mass. 

The mass loss, being a slow and surface effect, affects the total mass of the star along the RGB and the effective temperature, while it has a very small impact on the helium core mass and consequently on the Tip luminosity. 
For our reference set of models we adopted a Reimers mass loss scheme. For a star with a luminosity $L$, radius $R$ and total mass $M$, the mass loss rate is evaluated according to the following expression \citep{reimers75},
\begin{equation}
    \dot{M} = -4\times 10^{-13} \eta \frac{(L/\rm{L}_{\sun}) (R/\rm{R}_{\sun})}{(M/\rm{M}_{\sun})}\quad[\rm{M}_{\sun} \rm{yr}^{-1}]
\end{equation}
We used a parameter $\eta = 0.3$, which is similar to that used in the literature \citep{hidalgo18}, and is also compatible with that derived from the analysis of RGB stars \citep{miglio12}. When varying our reference input physics, we tested the impact of mass loss on the models by computing an additional set of models without mass loss, i.e setting $\eta =0$. It turns out that, a departure from $\eta = 0.3$ affects the LTRGB by less than $10^{-3}$~dex (see Table~\ref{tab:pert_eff}). 

\subsection{Initial chemical composition}
The initial chemical composition, and in particular, metallicity $Z$ and helium $Y$, affect both the evolutionary track of the star and the Tip luminosity. The helium content to be used to construct the models is obtained adopting the commonly used linear relation between helium abundance and metallicity as,
\begin{equation}
Y = Y_{\rm{p}} + \frac{\Delta Y}{\Delta Z} Z.
\end{equation}
For the primordial helium abundance we set $Y_p = 0.2470$ \citep{pitrou18,fields20}, and $\Delta Y/\Delta Z = 2$ \citep{tognelli21}.  To estimate the uncertainty on the adopted helium content, we used a conservative uncertainty on $Y_p$ of $\pm 0.0015$, which allows to take into account the maximum variation obtained by theoretical BBN calculations \citep[see e.g.][]{cyburt08,aver15,coc15,cyburt16,pitrou18,fields20}. For $\Delta Y/\Delta Z$, we assumed a conservative uncertainty of $\pm 1$ \citep[e.g.][]{casagrande07}. Combining the two uncertainties, we obtained the total error on the initial helium content, noticing that the uncertainty on $Y$ varies with the metallicity of the star, from $\delta Y \approx \pm 0.002$ at $Z=0.0001$ to  $\delta Y \approx \pm 0.02$ at $Z=0.020$. 

Instead of adopting only one value for the uncertainty $\delta Y$ for all metallicities, we computed the models for four initial helium values, which corresponds to a variation of the central helium of $1\sigma_Y,\,1.5\sigma_Y$ and $2\sigma_Y$. We adopted the following values for $\sigma_Y$:

\begin{align}
\sigma_Y = \left\{
\begin{array}{ll}
      0.002 & 0.0001 \le Z \le 0.001 \nonumber\\
      0.004 & 0.001 < Z \le 0.003 \nonumber\\
      0.006 & 0.003 < Z \le 0.010 \nonumber \\
      0.010 & 0.01 < Z \le 0.016 \nonumber \\
      0.015 & Z \gt 0.016 \nonumber \\
\end{array} 
\right. 
\end{align}

With the adopted values of $\sigma_Y$ we can estimate the effect of a relative variation of the initial helium content of about 2-4\% for $Z \lt 0.01$ and 8-10\% for $Z \ge 0.01$. An increase in the initial abundance of helium results in a hotter model, in any evolutionary stage. Consequently, helium burning is ignited at a lower luminosity. Table~\ref{tab:pert_eff} shows that a variation of $1\sigma_Y$ in the initial helium content does not produce a large effect, as it modifies the TRGB luminosity of about 2-3$\times 10^{-3}$~dex.

\subsection{Linear response of TRGB models} \label{sec:linearity}
\begin{figure}
\centering
\includegraphics[width=0.98\columnwidth]{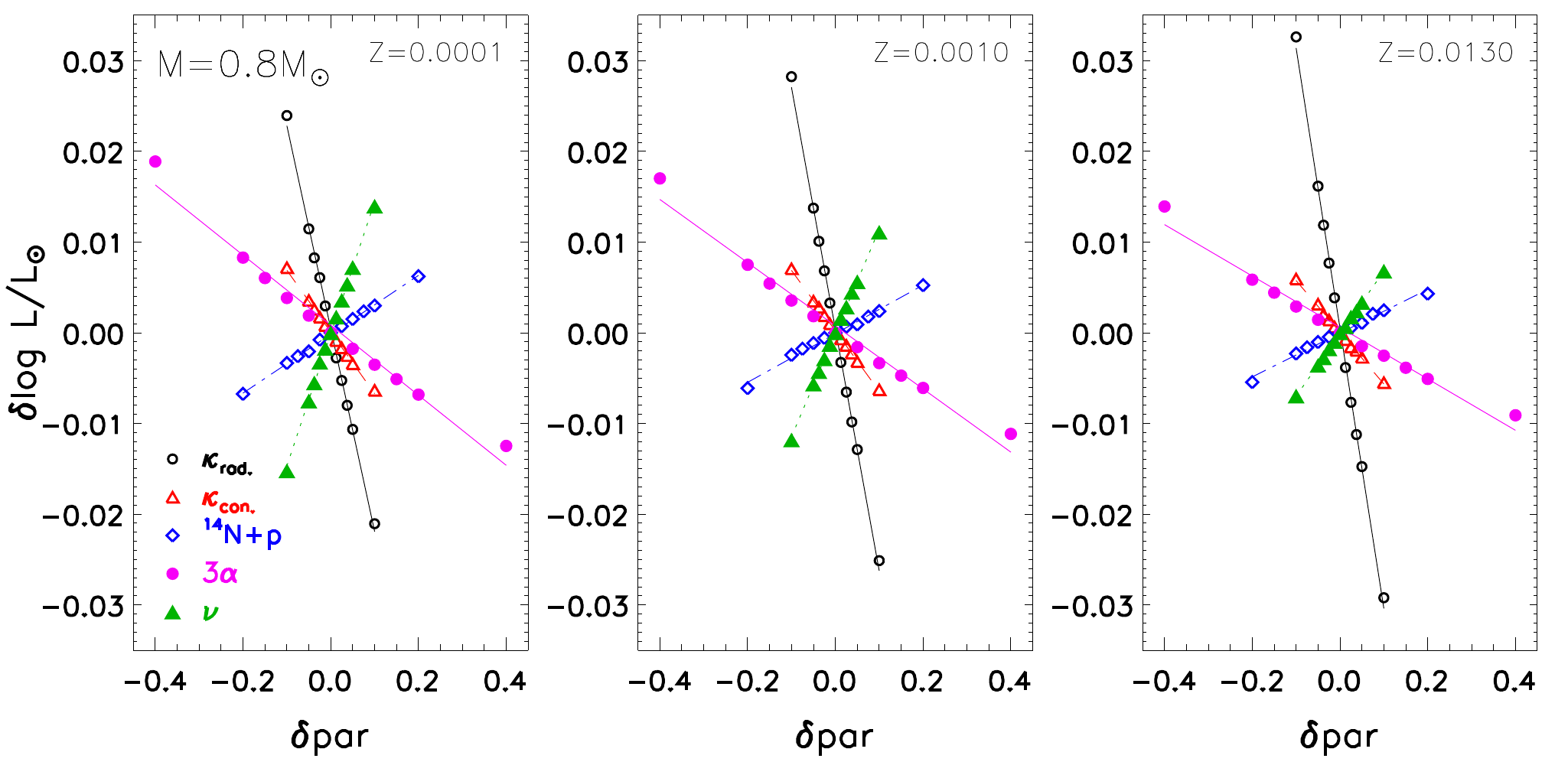}
\includegraphics[width=0.98\columnwidth]{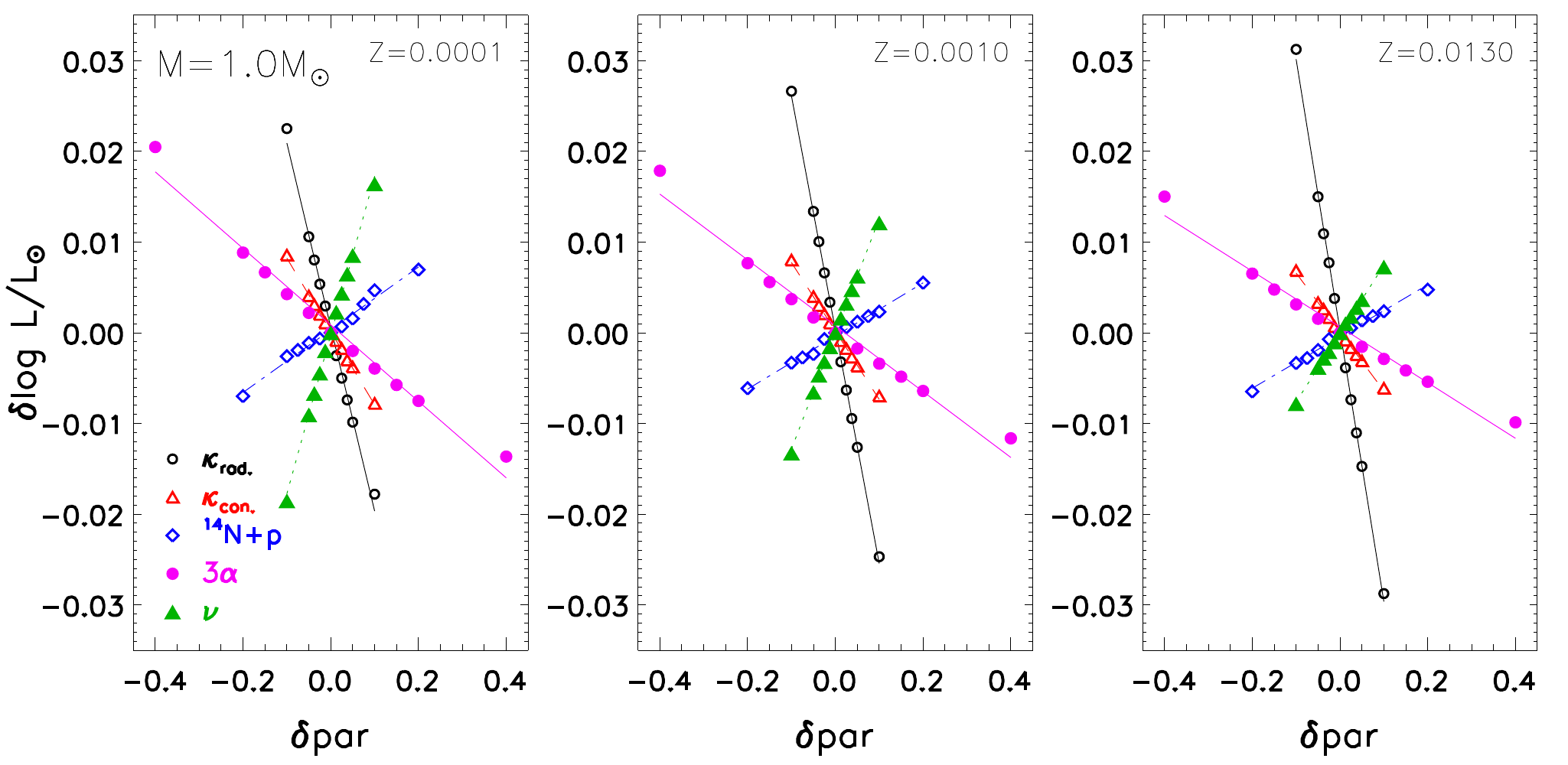}
\caption{Effect on the TRGB luminosity of the variation of \krad, \kcon, $3\alpha$, $^{14}$N(p,$\gamma$)$^{15}$O and $\nu$.  The models corresponds to perturbations of 0.25, 0.5, 0.75, 1 and 2$\sigma$ for each considered quantity. The linear fit to the luminosity vs perturbation is overplotted (see text). Top panels: Effects on a 0.8~\msun{} for three chemical compositions, namely (Z,Y) = (0.0001, 0.2472), (0.0010, 0.2490) and (0.0130, 0.2730). Bottom panels: Same as top panels, but for $M=$1.0~\msun.}
\label{fig:lin_fit}
\end{figure}
So far, we analysed the impact of each quantity on the Tip luminosity due to the perturbation of each parameter/input physics by 1$\sigma$, at least for those cases where the error of the considered quantity can be defined. Here, we will ask the question of whether the Tip model responds linearly to a small perturbation of each quantity. In this regard, \citet{valle13} showed that Tip models respond linearly to small perturbations of a certain set of quantities. We extend their analysis considering all the perturbed quantities and in a wider metallicity range. To do this, we computed models for individual perturbation of each quantity, but instead of varying it by $1\sigma$, we varied it by 0.25, 0.5, 0.75 and $2\sigma$ respectively. For each case we obtained the Tip model and compared it to the reference set. Figure~\ref{fig:lin_fit} shows the variation of the Tip luminosity, $\delta \log L/$\lsun$ = \log L/$\lsun(perturbed)$ - \log L/$\lsun(reference), against the corresponding perturbation $\delta G$, for $1 M_{\odot}$ and three metallicities (Z = 0.0001, 0.0010, 0.0130), only for the quantities that mainly affect the Tip luminosity. These are \krad, \kcon, $3\alpha$, $^{14}$N+p and $\nu$. In Figure \ref{fig:lin_fit} we also over-plotted the linear fit $\delta \log L/$\lsun$ = |_{G_j}=\beta_{G_j} \delta_{G_j}/G_j + \alpha_{G_j}$, with the corresponding coefficients for the fit listed in Table~\ref{tab:fit}. In all selected cases, the models respond linearly to the perturbation, modulo small numerical fluctuations, while the results are weakly dependent on the metallicity. We conclude that the hypothesis of linear response to the perturbed quantities holds for the Tip model. We will get back to this issue later on, to compare against the results of our Monte-Carlo analysis.
\begin{table}
\centering
\caption{Coefficients for the linear fit of the TRGB luminosity vs parameters perturbation ($\delta_j$), namely $\delta \log L/$\lsun{}$|_{G_j}=\beta_{G_j} \delta_{G_j}/G_j$ for the same quantities shown in Fig.~\ref{fig:lin_fit}. The coefficients have been estimated  fitting the luminosity variation at fixed $Z$ for all the models in the mass range [0.8,1.2]~\msun.} 

\begin{tabular}{lrrr}
\hline
 & \multicolumn{3}{c}{linear fit parameters}\\
 & Z=0.0001  & Z=0.0010 & Z=0.0130 \\
 $G_j$ & $\beta_{G_j}$ & $\beta_{G_j}$ & $\beta_{G_j}$ \\
\hline
\hline
$\kappa_{\rm{rad.}}$ & $-1.99\times 10^{-1}$ & $-2.45\times 10^{-1}$ & $-2.98\times 10^{-1}$ \\
$\kappa_{\rm{con.}}$ & $-7.38\times 10^{-2}$ & $-6.92\times 10^{-2}$ & $-6.12\times 10^{-2}$ \\
$^{14}$N(p,$\gamma)^{15}$O & $ 3.24\times 10^{-2}$ & $ 2.88\times 10^{-2}$ & $ 2.33\times 10^{-2}$ \\
$\alpha(\alpha\alpha,\gamma)^{12}$C & $-3.68\times 10^{-2}$ & $-3.20\times 10^{-2}$ & $-2.70\times 10^{-2}$ \\
$\nu$ & $ 1.65\times 10^{-1}$ & $ 1.23\times 10^{-1}$ & $ 7.16\times 10^{-2}$ \\
\hline
\end{tabular}
\label{tab:fit}
\end{table}

\begin{figure*}
\centering
\includegraphics[width=0.32\linewidth]{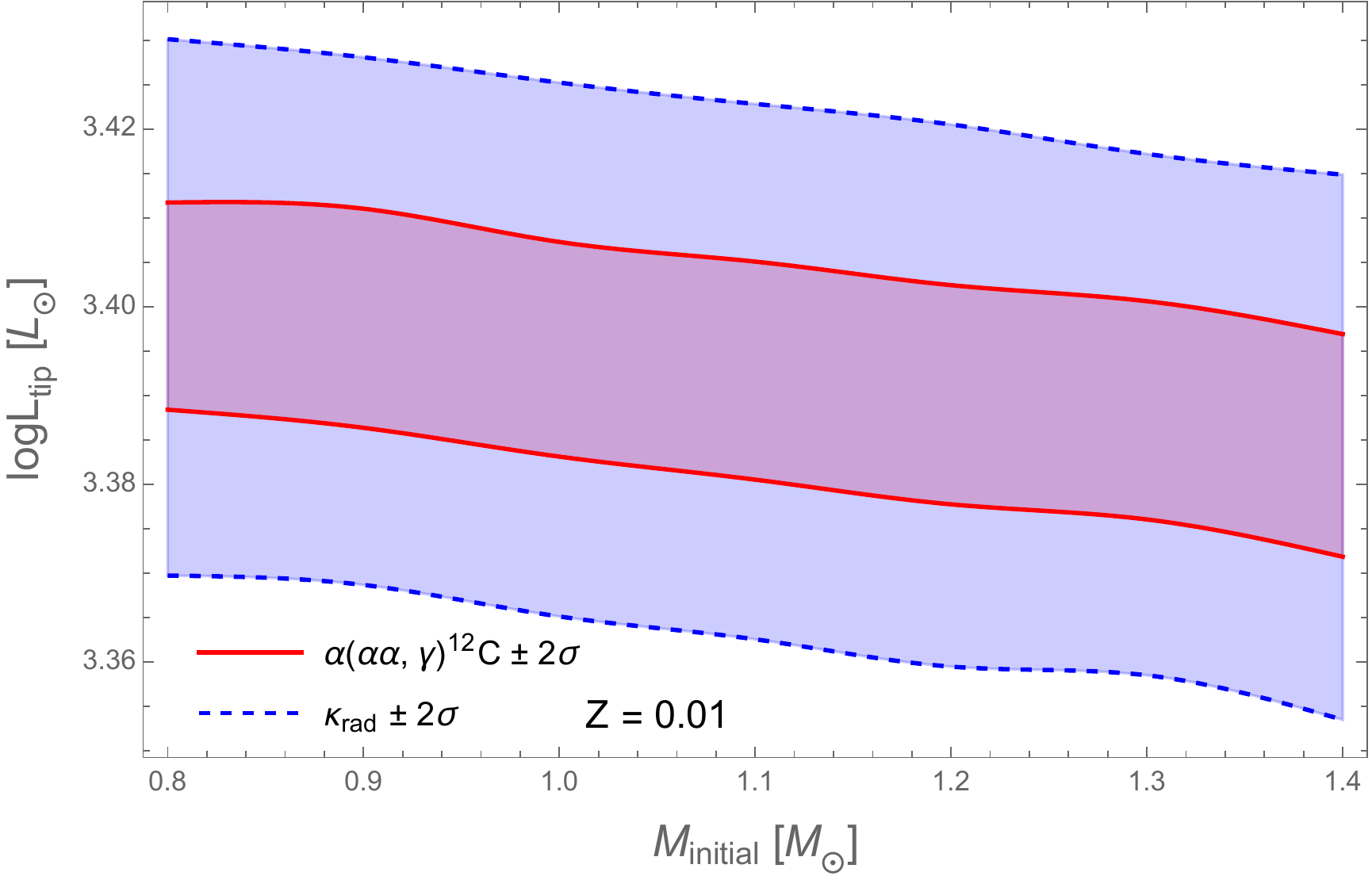}
\includegraphics[width=0.32\linewidth]{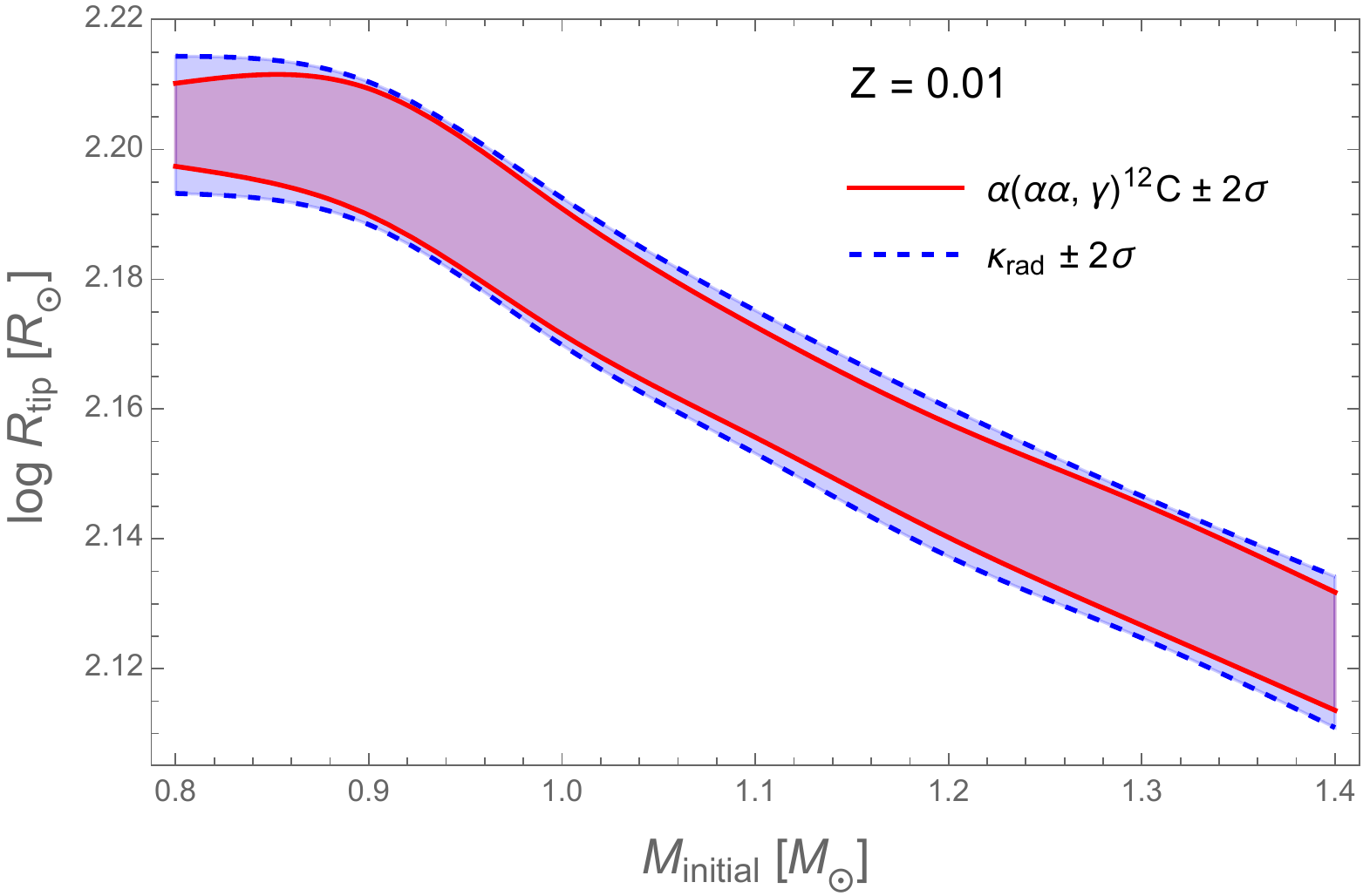}
\includegraphics[width=0.32\linewidth]{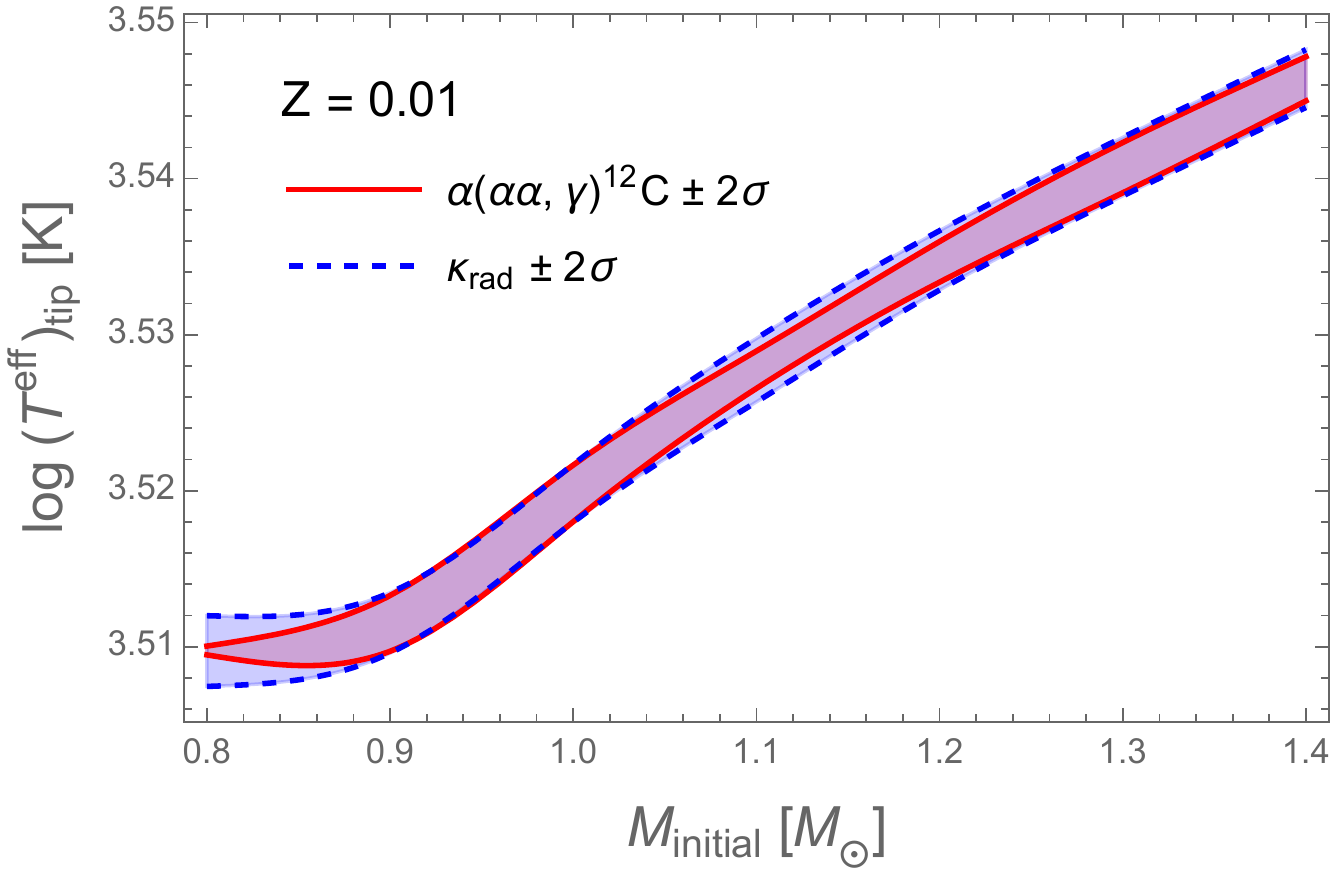}
\includegraphics[width=0.32\linewidth]{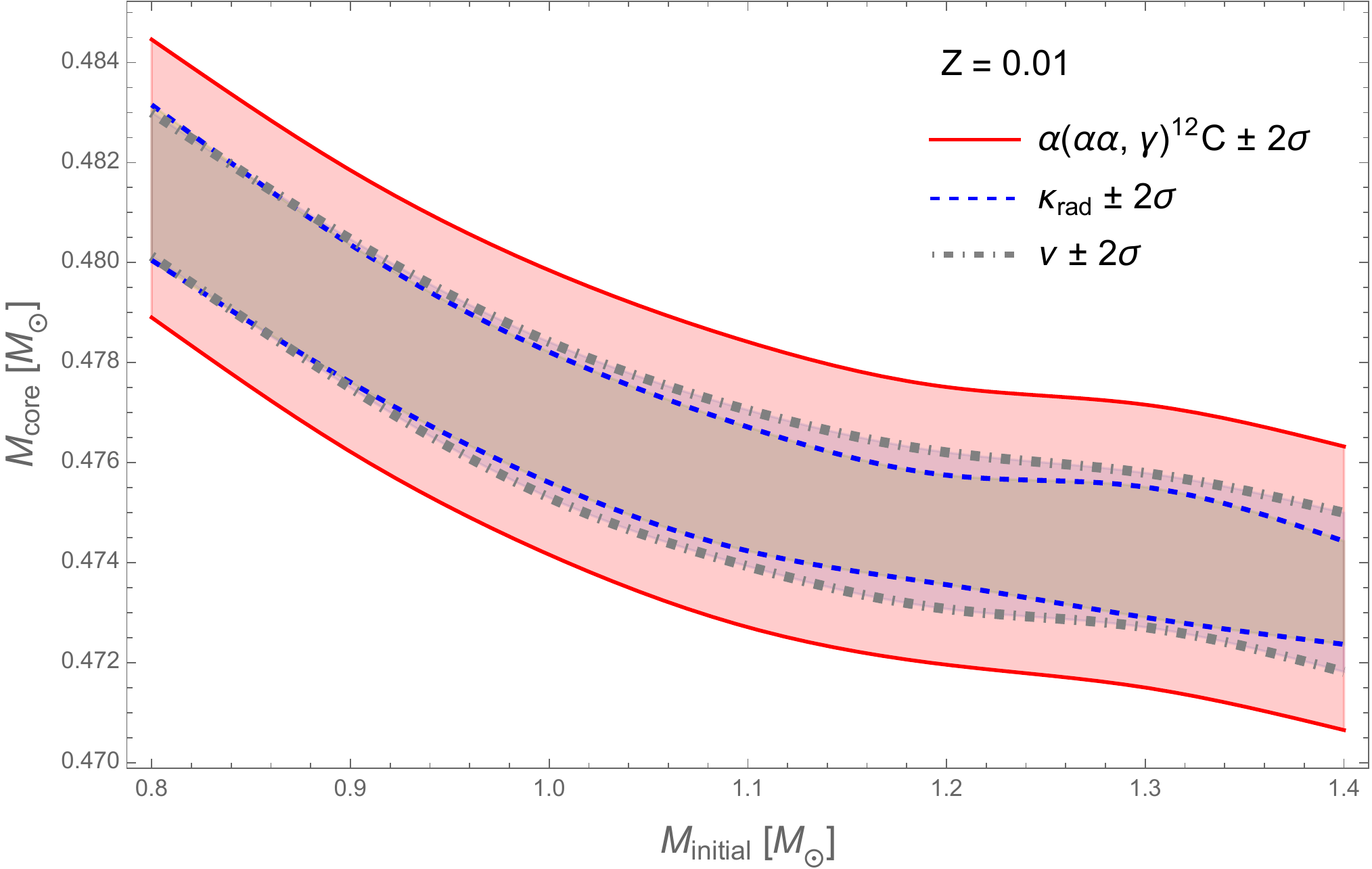}
\includegraphics[width=0.32\linewidth]{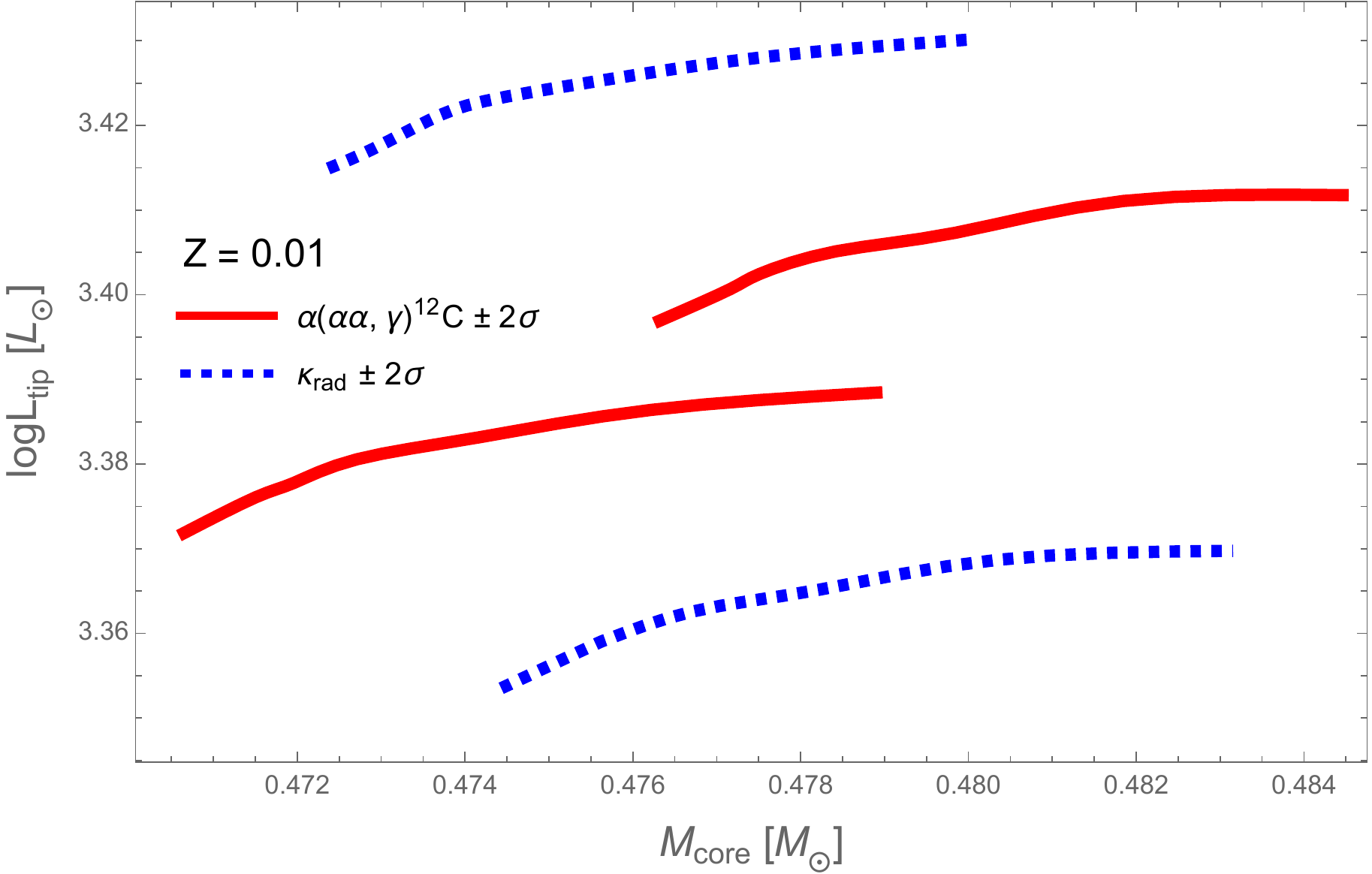}
\includegraphics[width=0.32\linewidth]{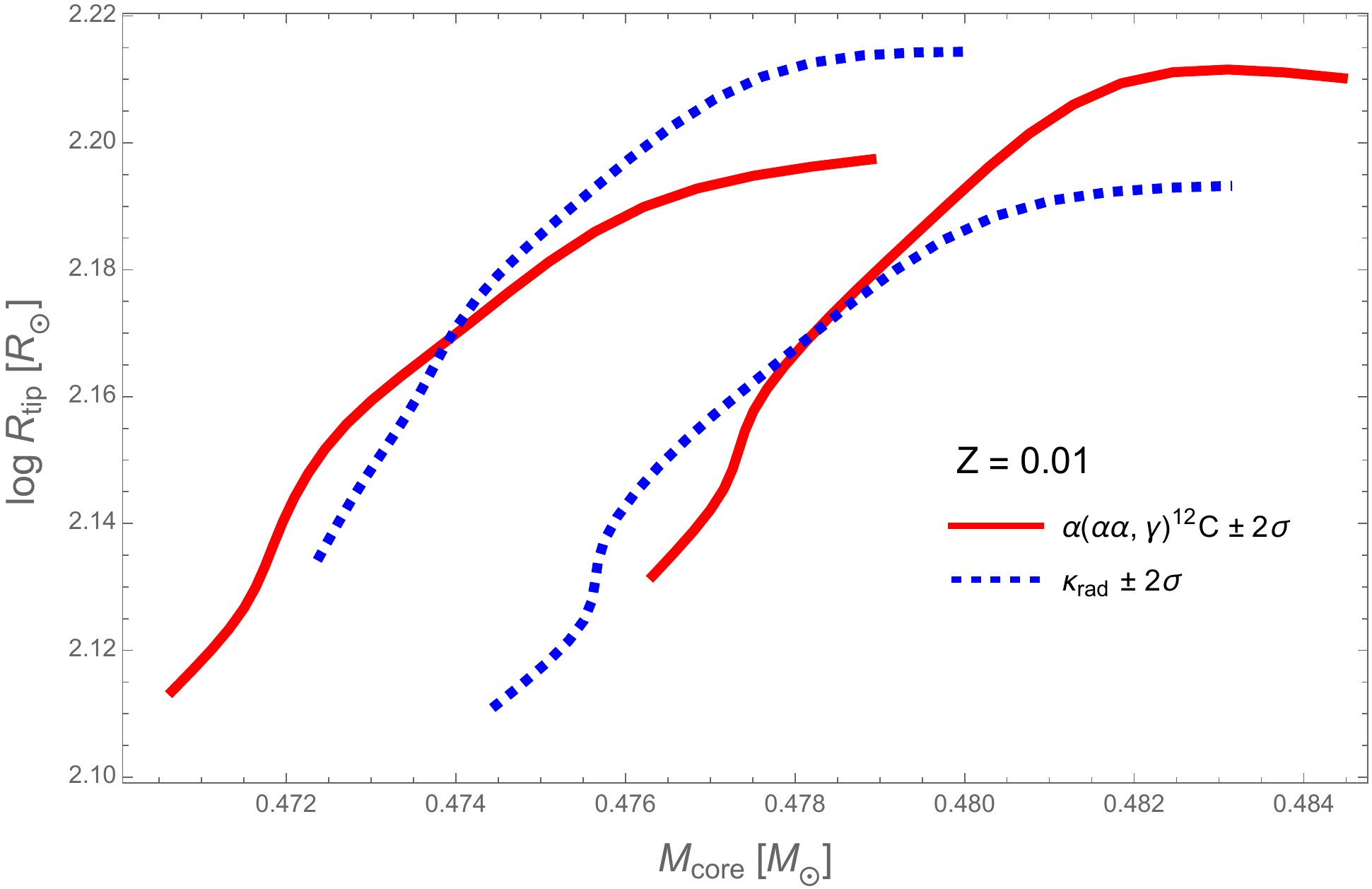}
\caption{ A sample illustration of the variation of key quantities at Tip luminosity under the uncertainty in the value of radiative opacity, the $3\alpha$ reaction and neutrino emission. As can be seen, the uncertainty of radiative opacity has the dominant effect on the luminosity on the Tip, whereas it is comparable with that of the $3\alpha$ reaction for radius and effective temperature. The picture is similar for different values of metallicity Z (not shown). In the second and third plot of the second panel, one notices the relatively mild dependence of the TRGB luminosity with the mass of the degenerate core $M_{c}$, as expected from theoretical arguments.  Notice also that, in the latter plots, the offset in the respective curves along the horizontal axis is due to the different starting value acquired by the core mass $M_{c}$ for different values of the initial mass $M_{\rm{initial}}$.  }
\label{fig:gauss_total}
\end{figure*}

\begin{figure*}
\centering
\includegraphics[width=0.32\linewidth]{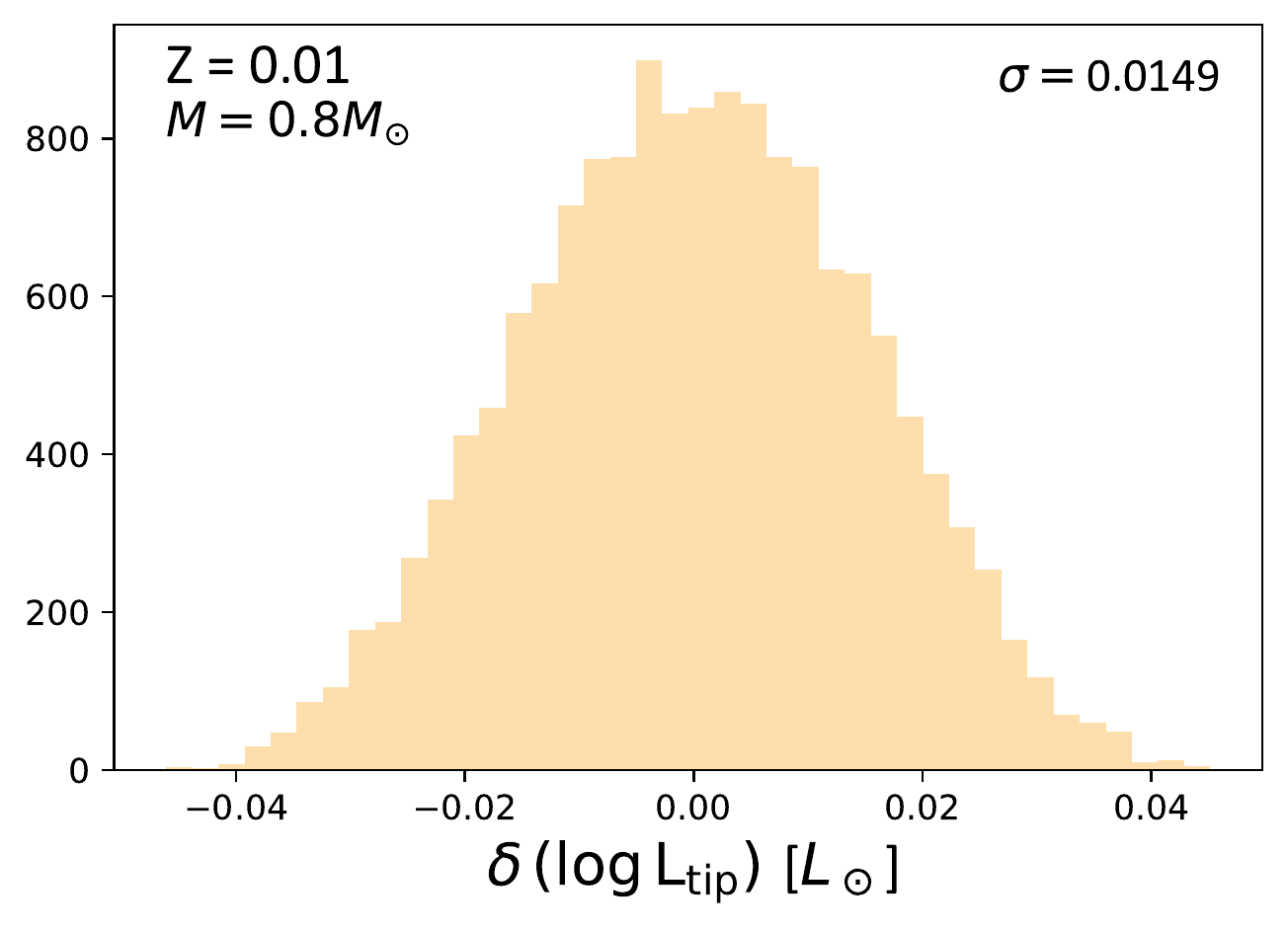}
\includegraphics[width=0.32\linewidth]{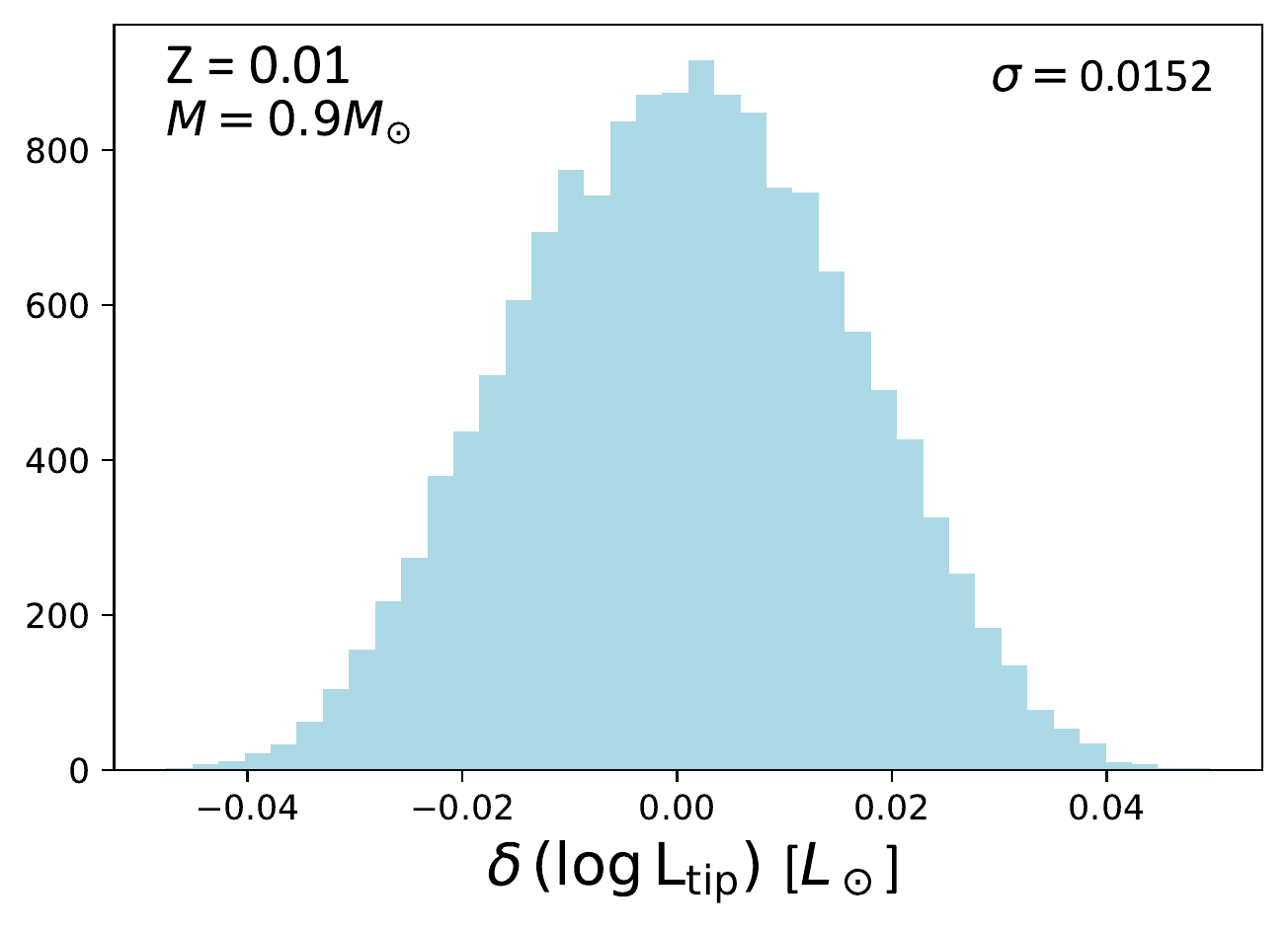}
\includegraphics[width=0.32\linewidth]{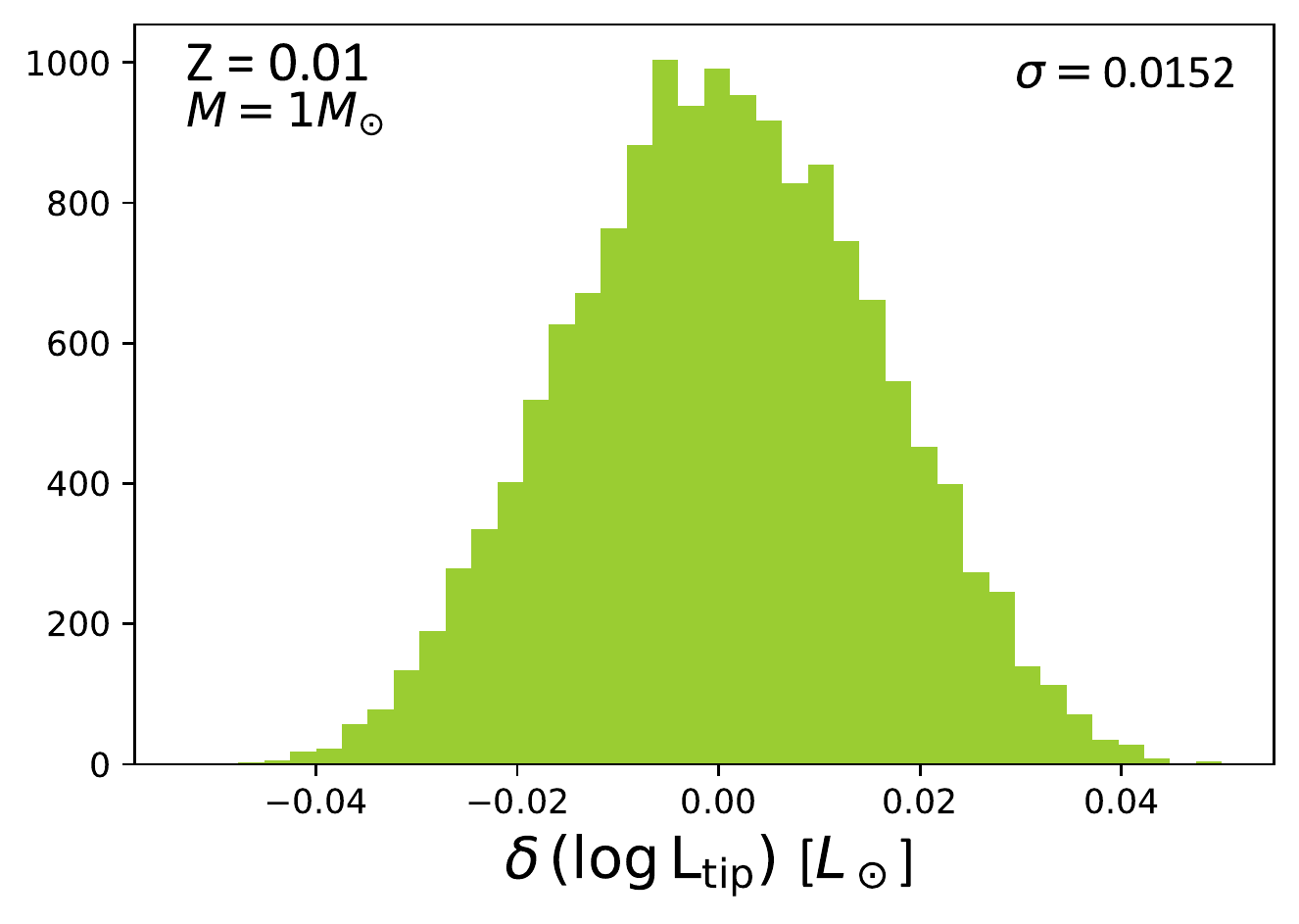}
\includegraphics[width=0.32\linewidth]{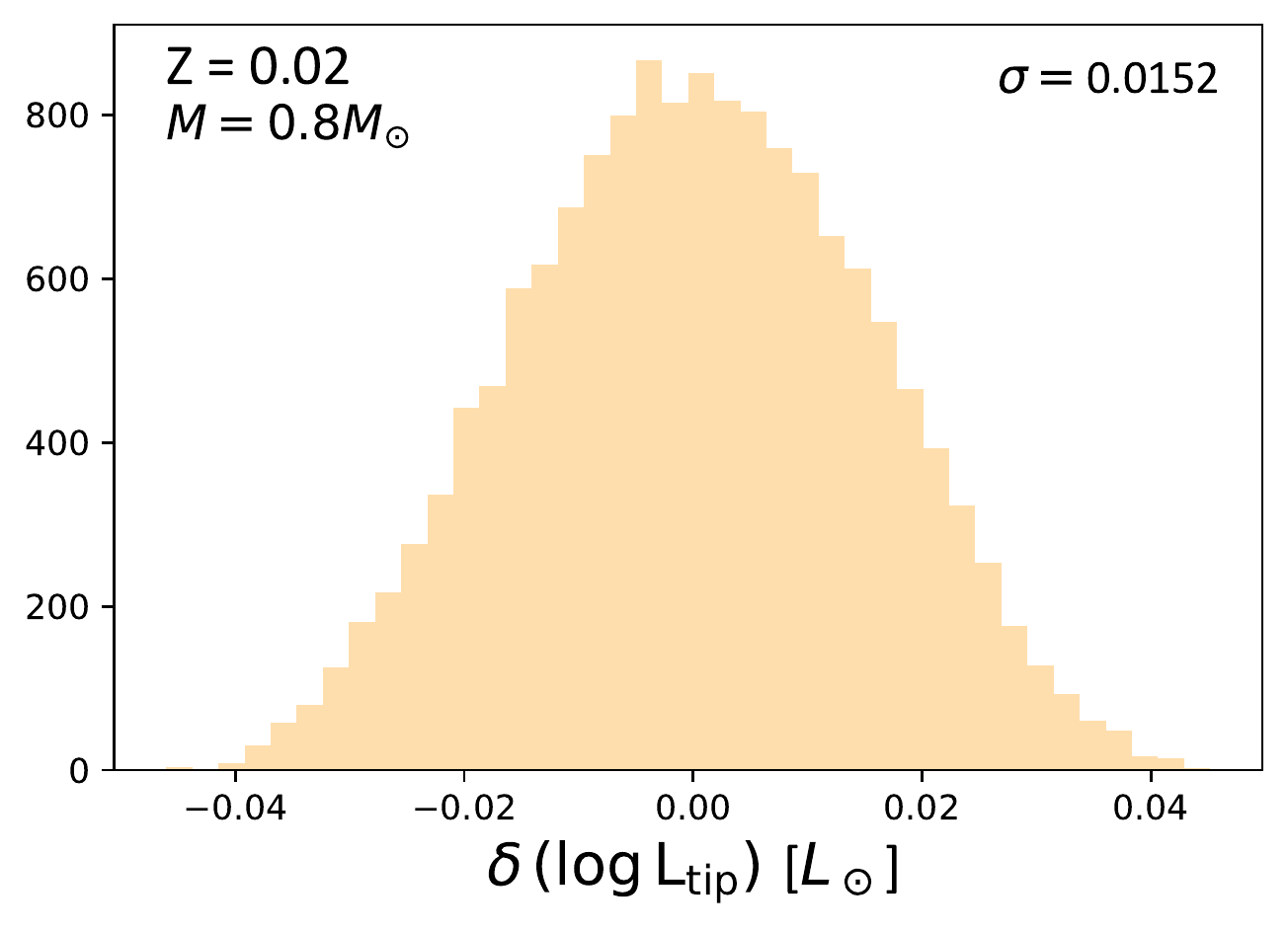}
\includegraphics[width=0.32\linewidth]{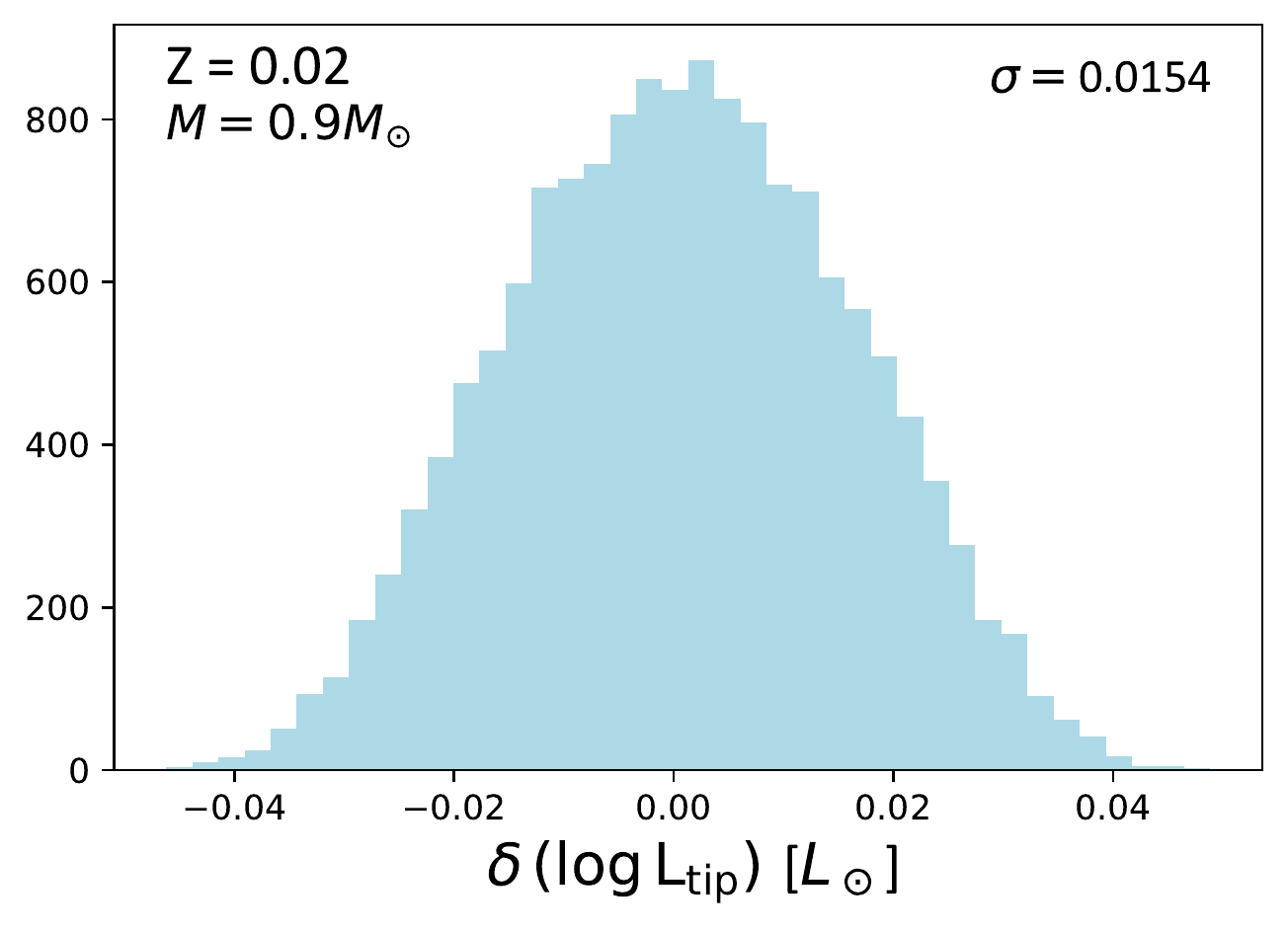}
\includegraphics[width=0.32\linewidth]{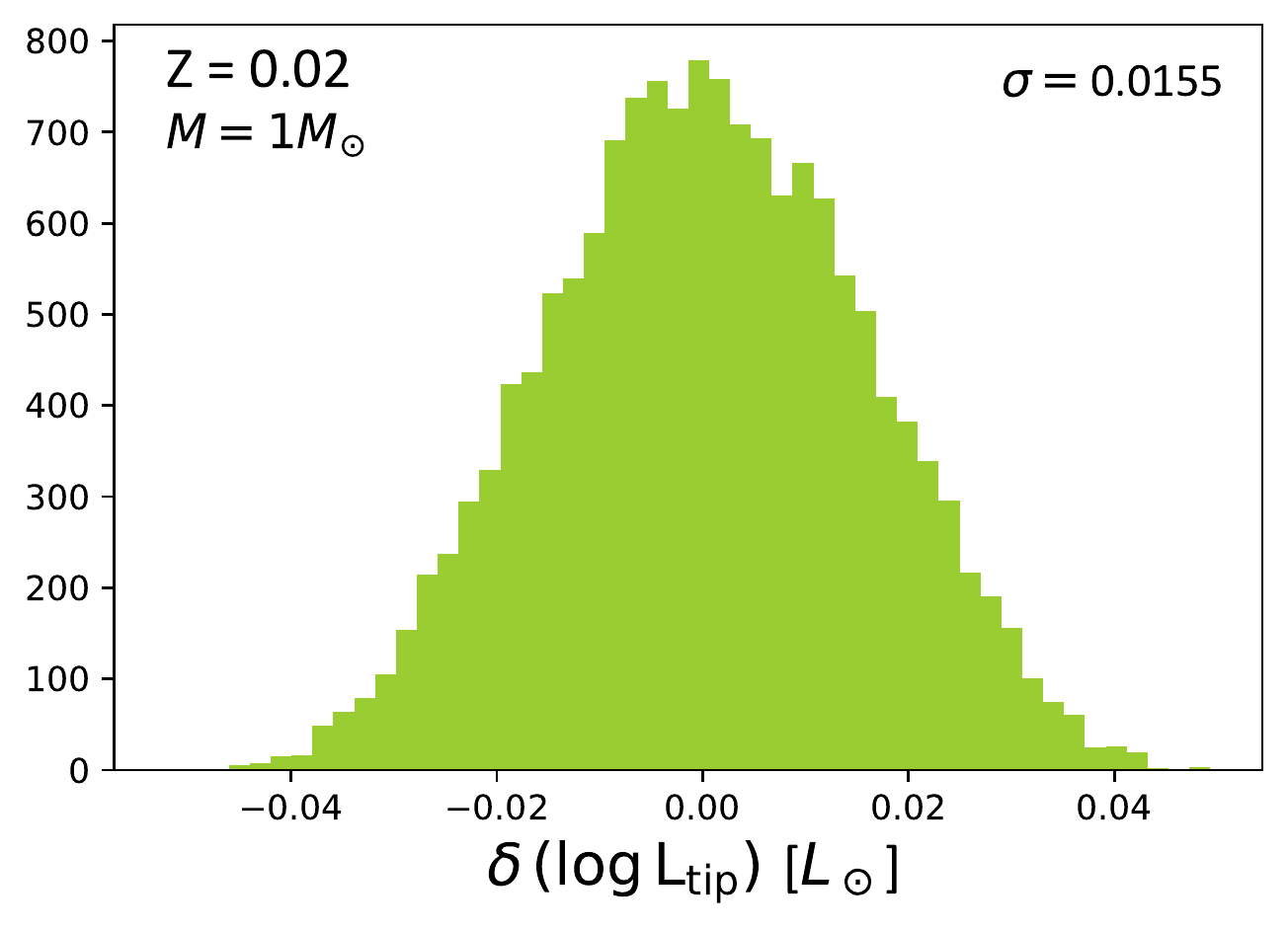}

\caption{ Histograms for the predicted value of the TRGB luminosity, resulting from the {\it simultaneous} variation of radiative opacity (\krad), conductive opacity (\kcon), the reactions $^{14}$N(p,$\gamma)^{15}$O and $\alpha(\alpha\alpha,\gamma)^{12}$C as well as neutrino emission ($\nu$). The value of initial helium abundance is held fixed at $Y = 0.247 $ for $Z = 0.01$, and $Y = 0.257$ for $Z = 0.02$. Each row of plots corresponds to a different choice of initial mass, $0.8, 0.9, 1.0 M_{\odot}$. The variation of each observable is in the sense $\delta(\rm{observable}) = (\rm{reference \; model}) - (\rm{modified \; model})$, with the modified model corresponding to an evaluation of the Tip luminosity from a randomly chosen value for each input physics parameter within the interval $[-2\sigma, +2\sigma]$. The sampling is based on $N \simeq 15,000$ points sampled from an interpolated version of the grid of models, assuming each input physics parameter is Gaussianly distributed with a mean at its reference value, and a respective $\sigma$ value according to Table \ref{tab:pert_eff}. As expected, the uncertainty on the luminosity increases with metallicity at constant initial mass. The results shown here should be compared to those presented in Table \ref{tab:ind_tot}.} \label{fig:gauss_total_L}
\end{figure*}

\begin{figure*}
\centering
\includegraphics[width=0.32\linewidth]{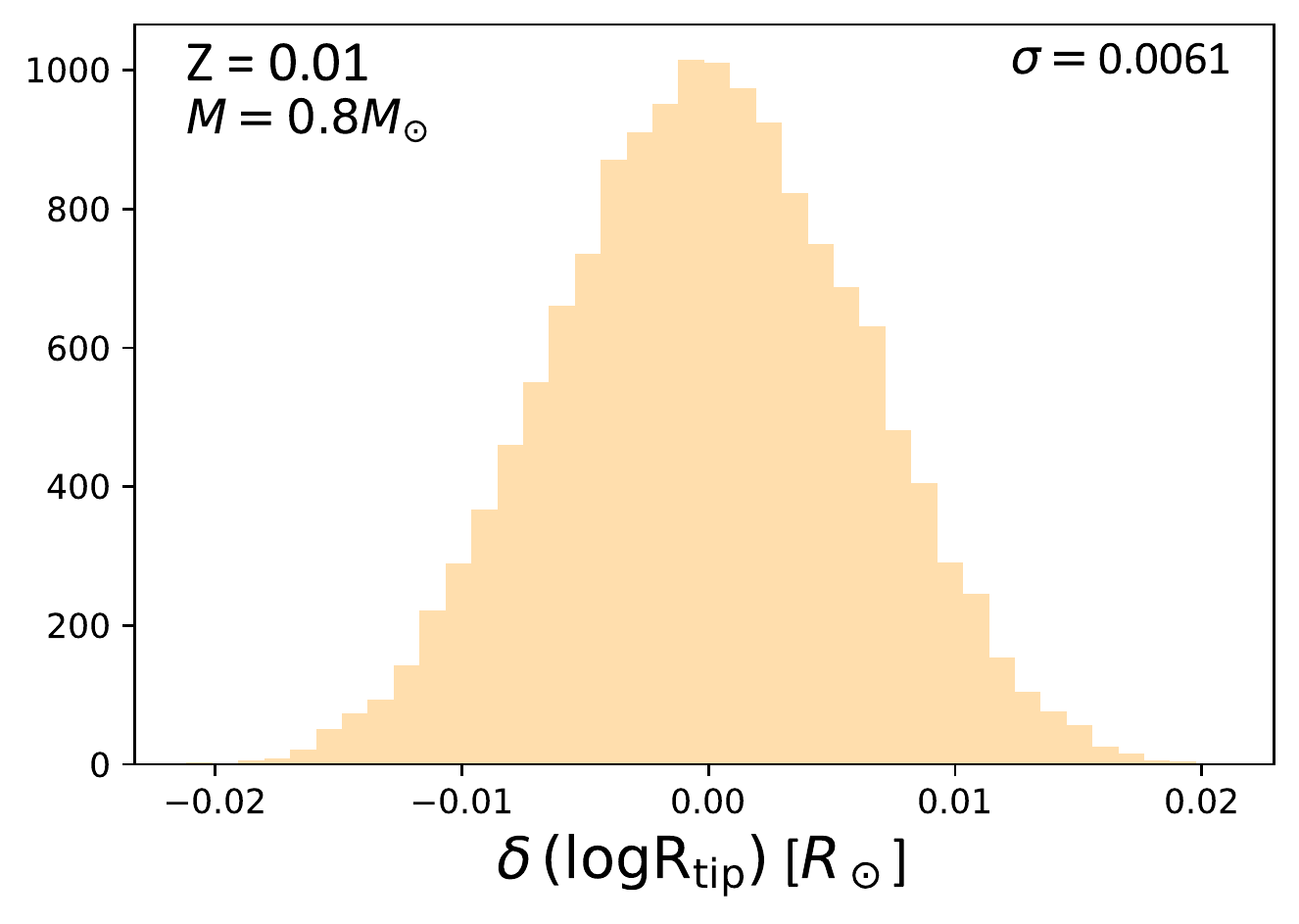}
\includegraphics[width=0.32\linewidth]{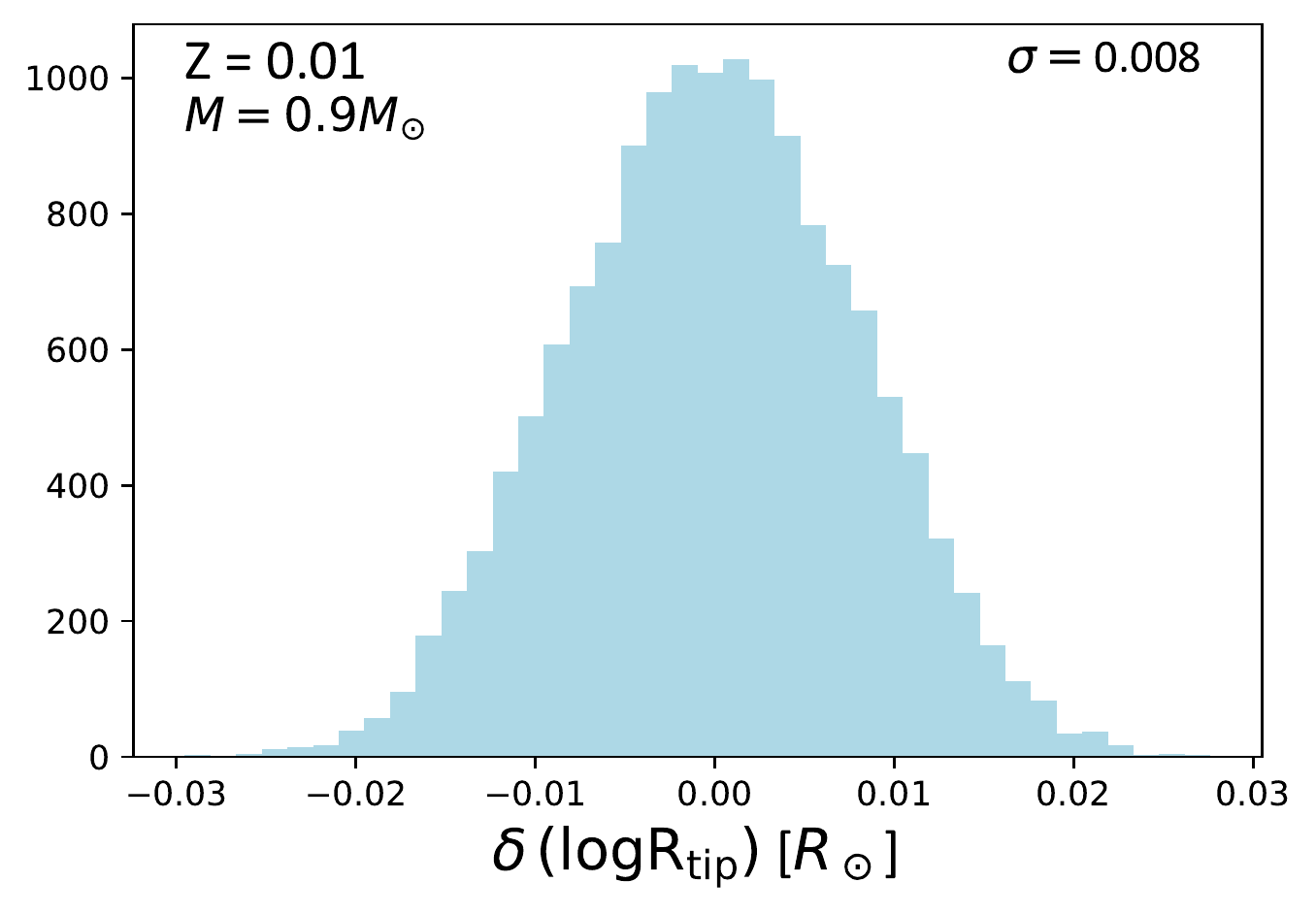}
\includegraphics[width=0.32\linewidth]{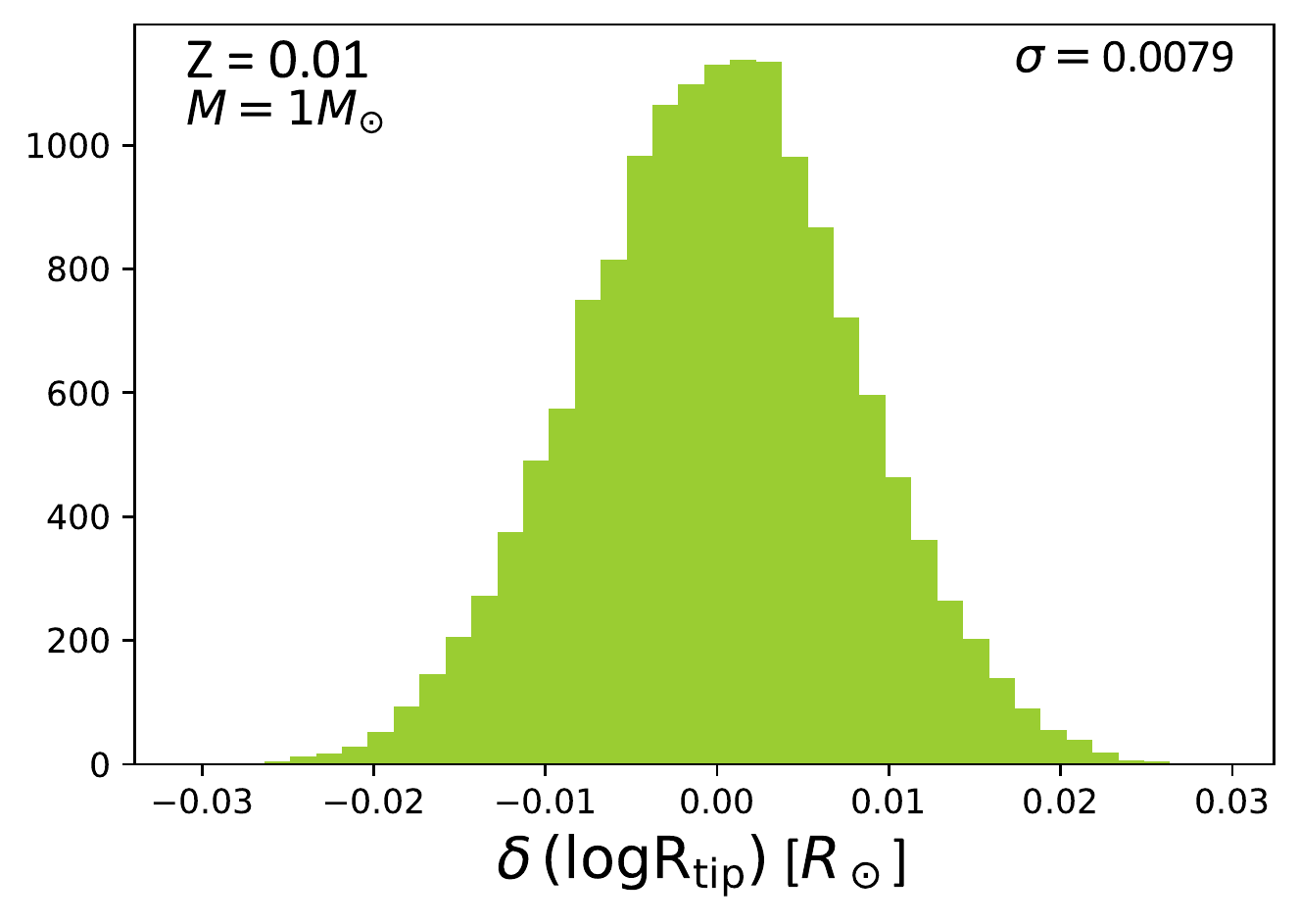}
\includegraphics[width=0.32\linewidth]{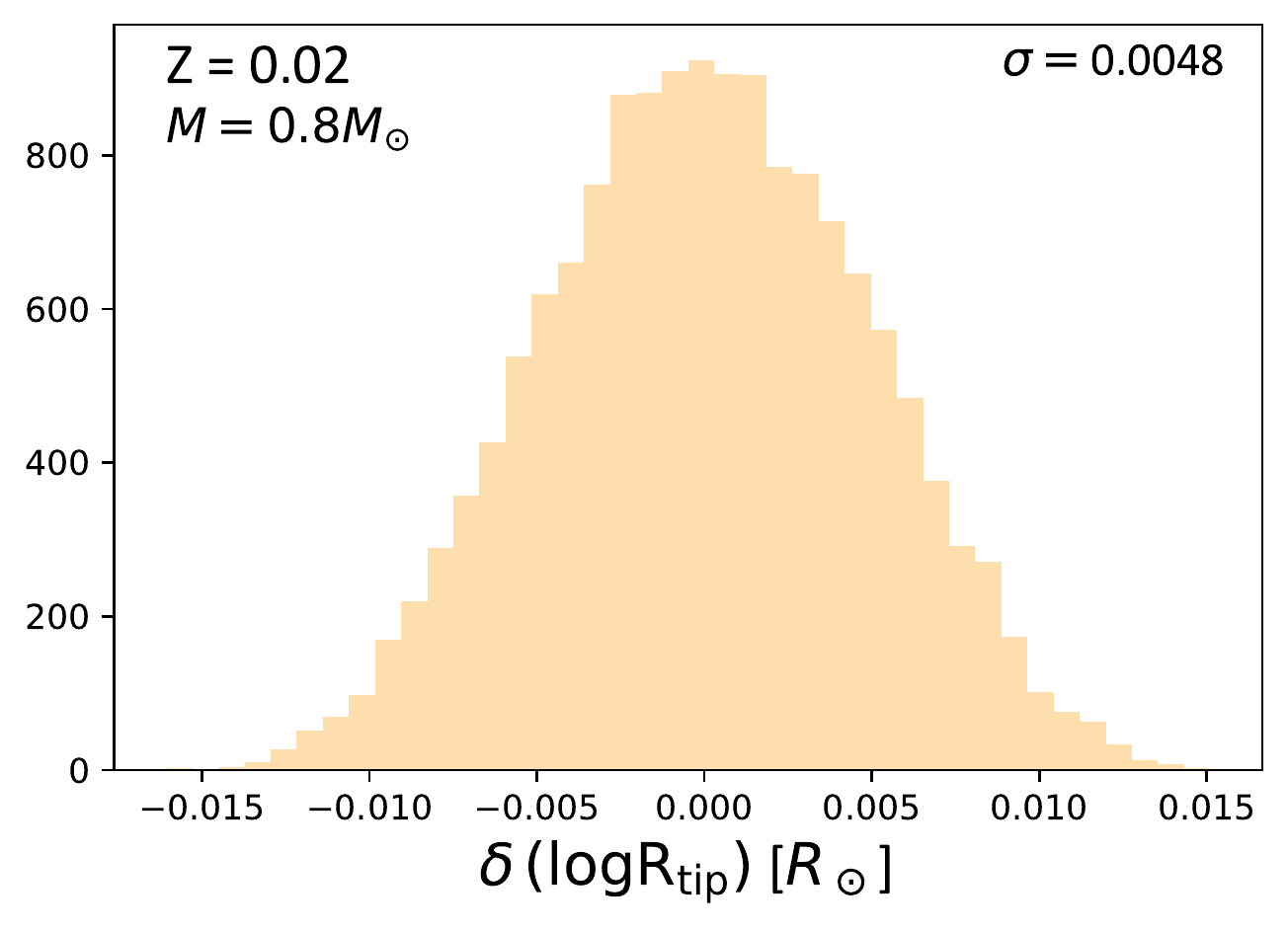}
\includegraphics[width=0.32\linewidth]{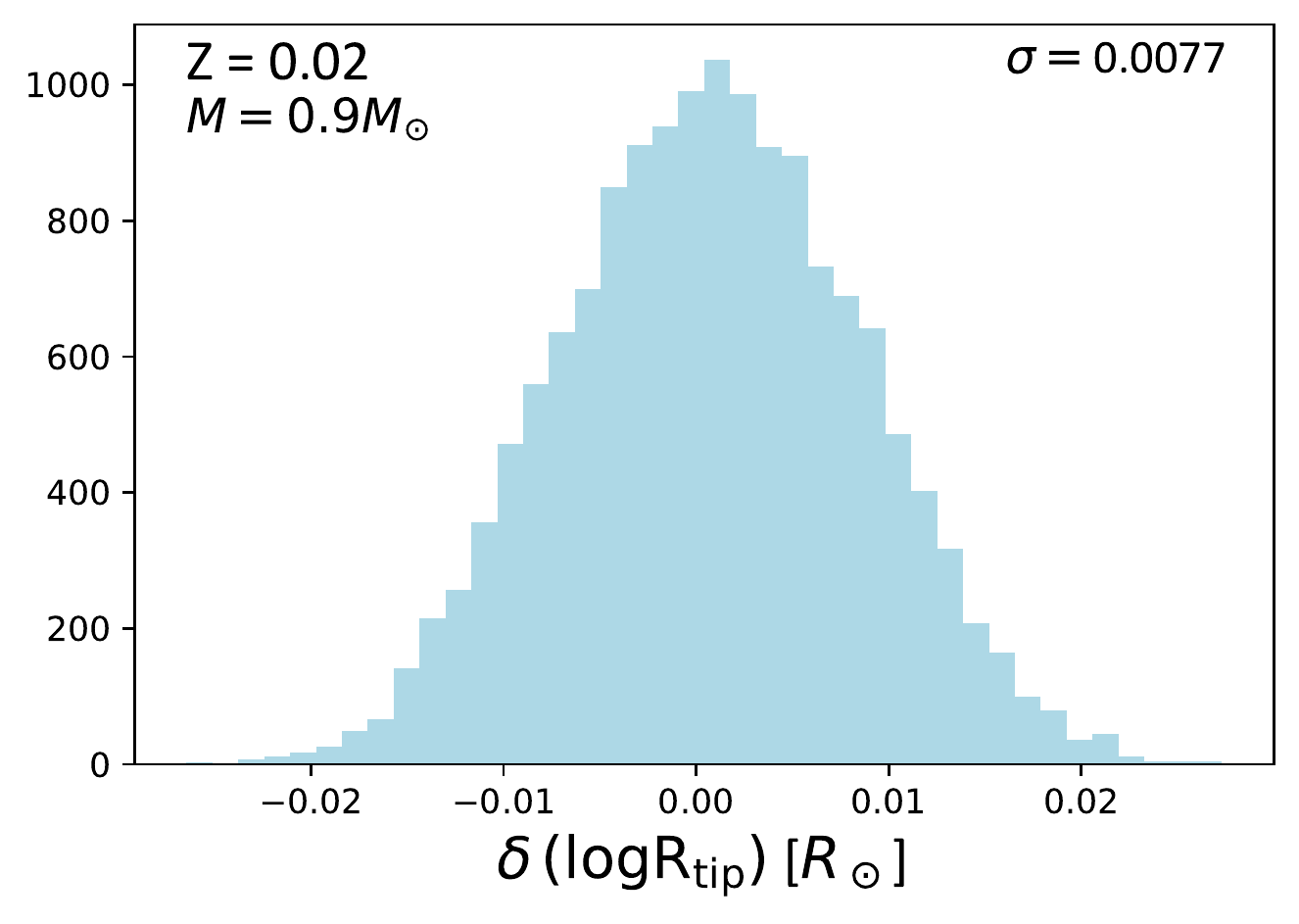}
\includegraphics[width=0.32\linewidth]{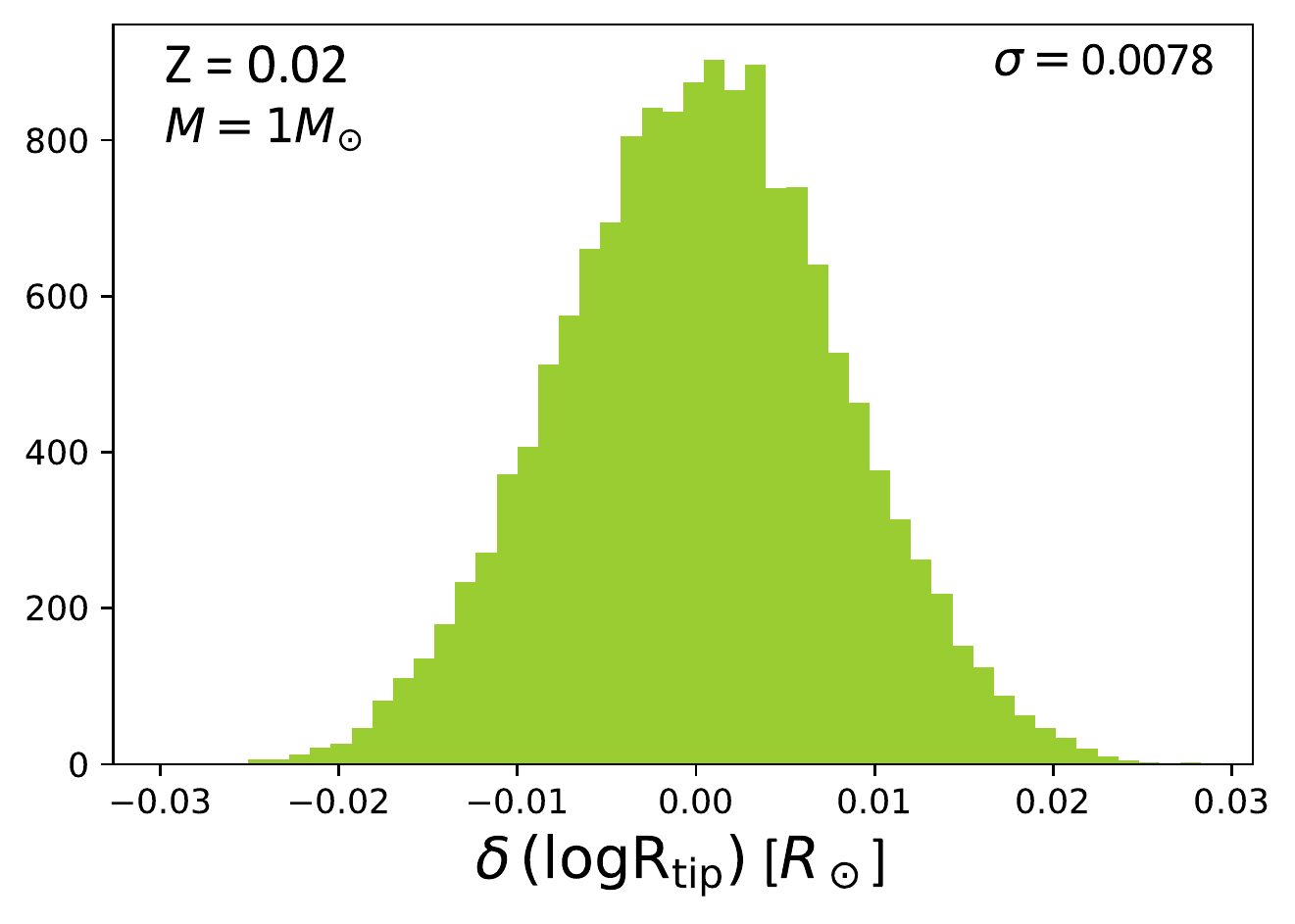}
\caption{Same as Figure \ref{fig:gauss_total}, but for the total stellar radius at Tip luminosity.} \label{fig:gauss_total_R}
\end{figure*}

\begin{figure*}
\centering
\includegraphics[width=0.32\linewidth]{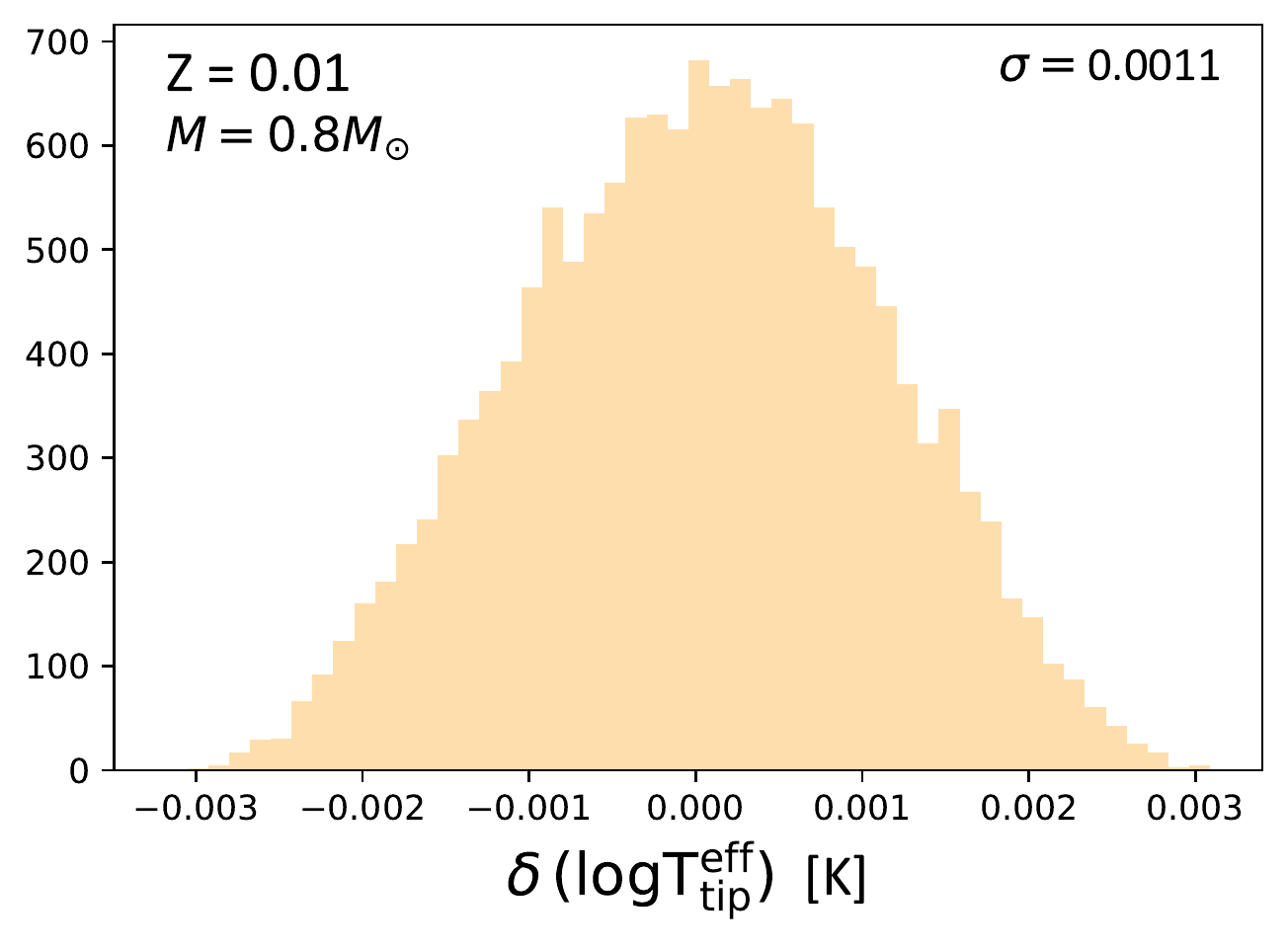}
\includegraphics[width=0.32\linewidth]{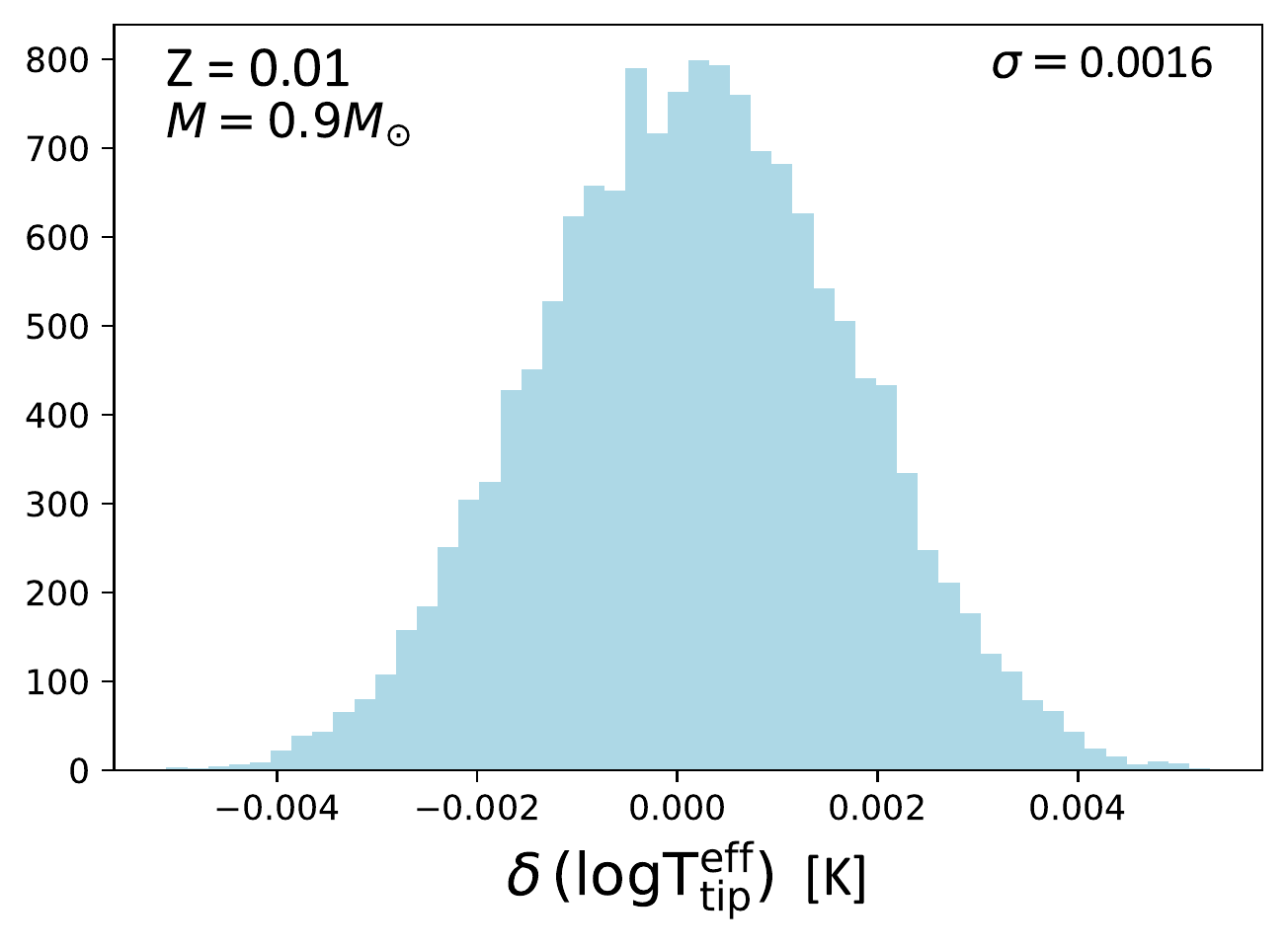}
\includegraphics[width=0.32\linewidth]{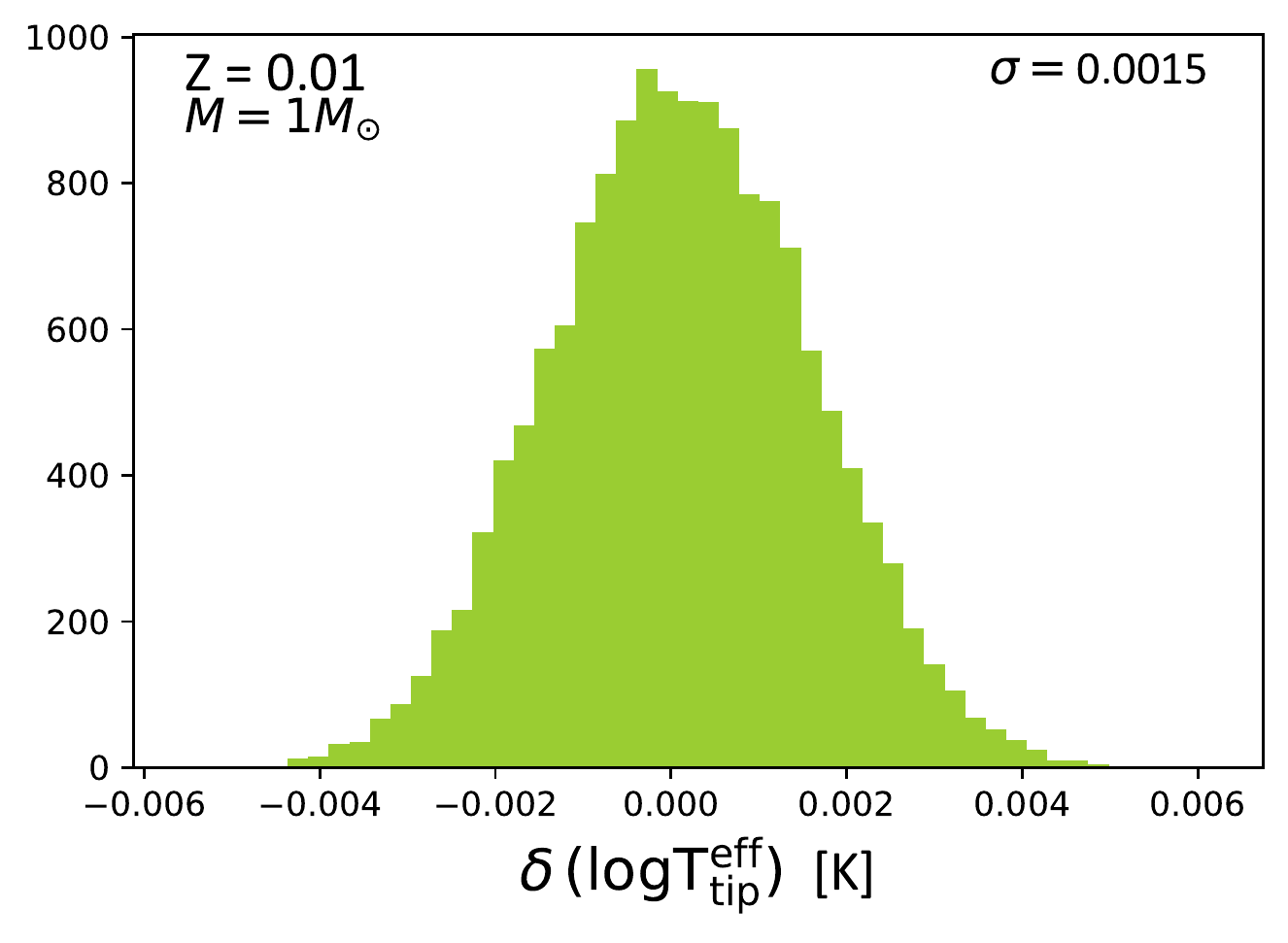}
\includegraphics[width=0.32\linewidth]{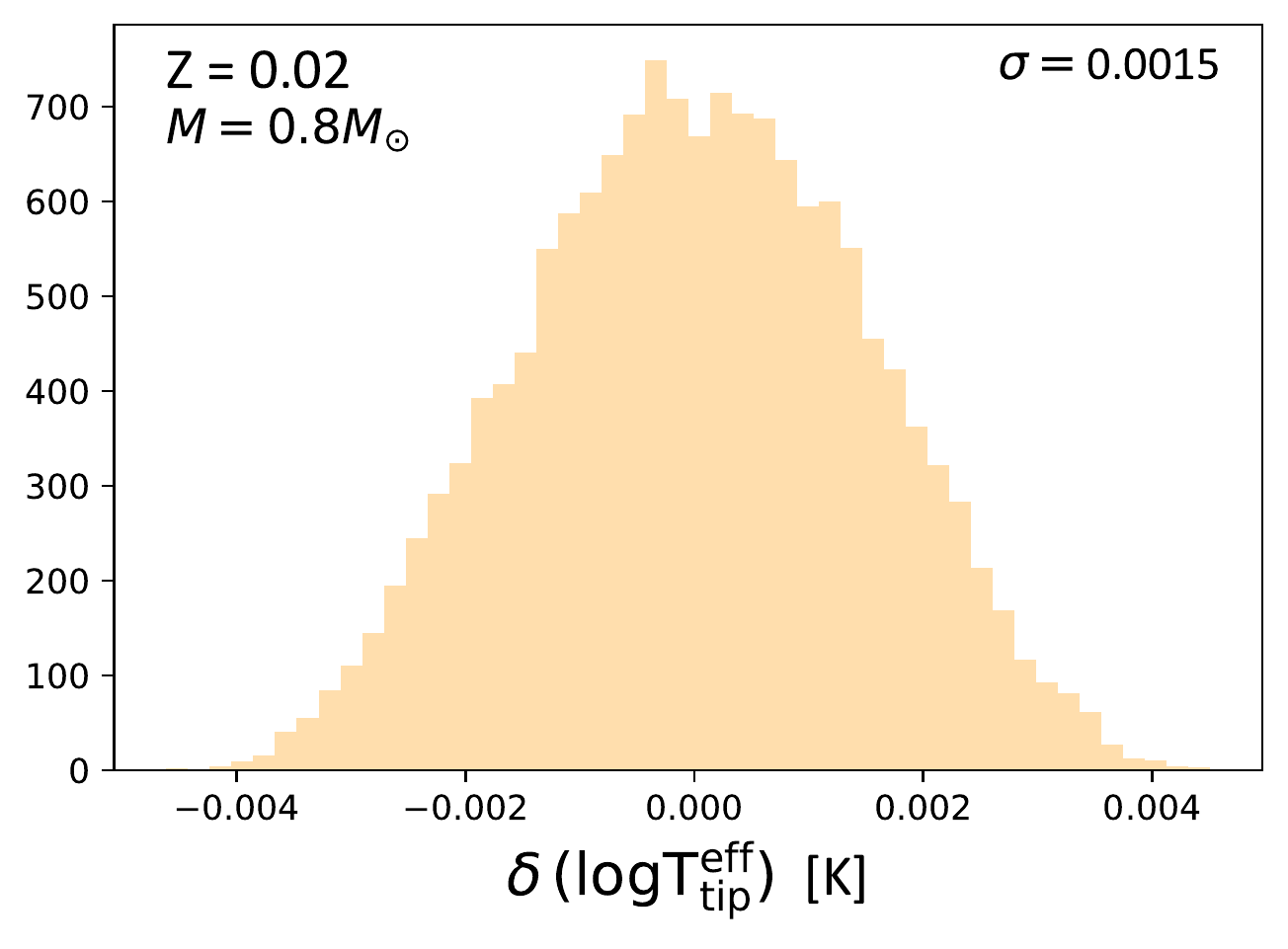}
\includegraphics[width=0.32\linewidth]{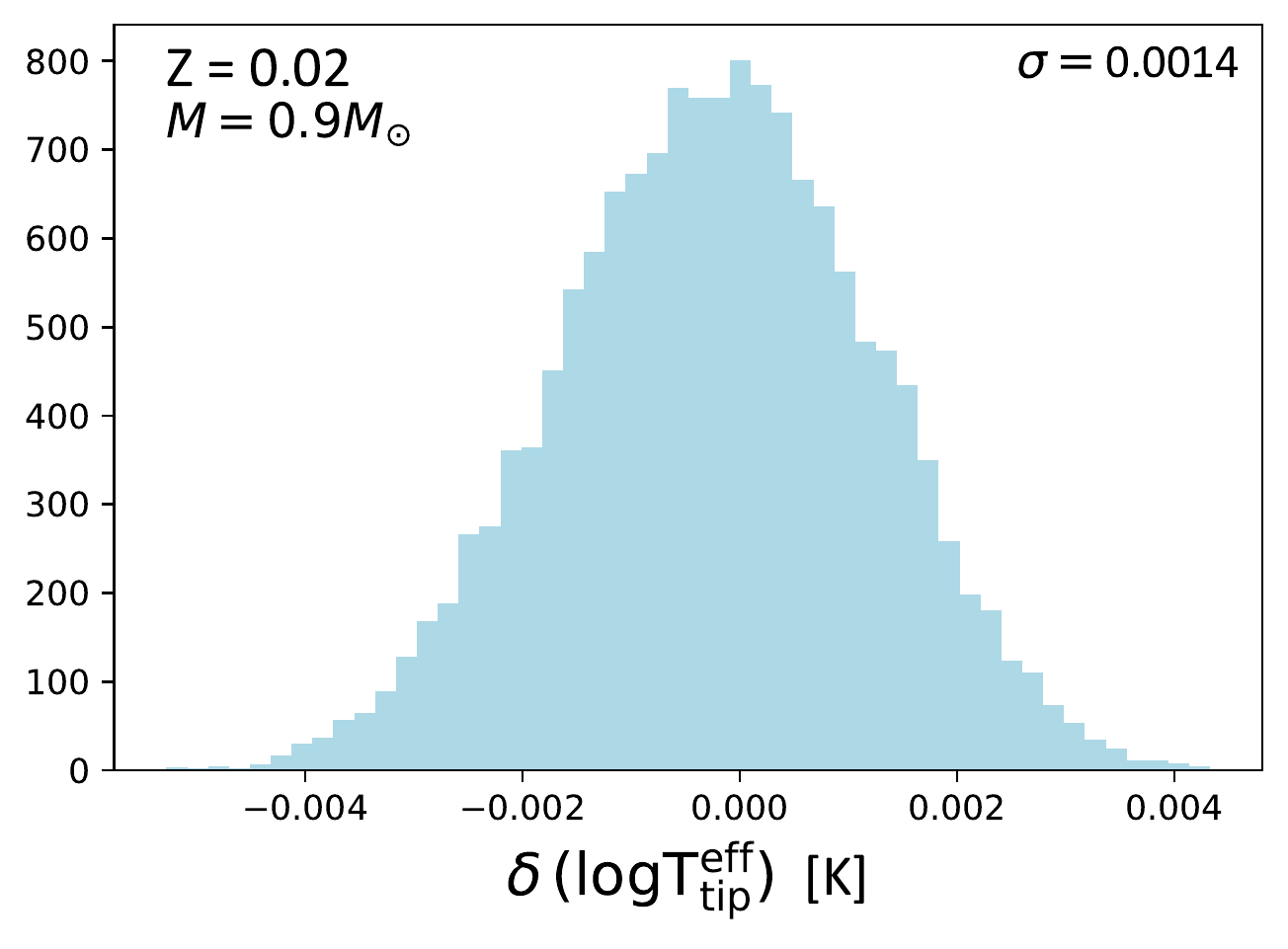}
\includegraphics[width=0.32\linewidth]{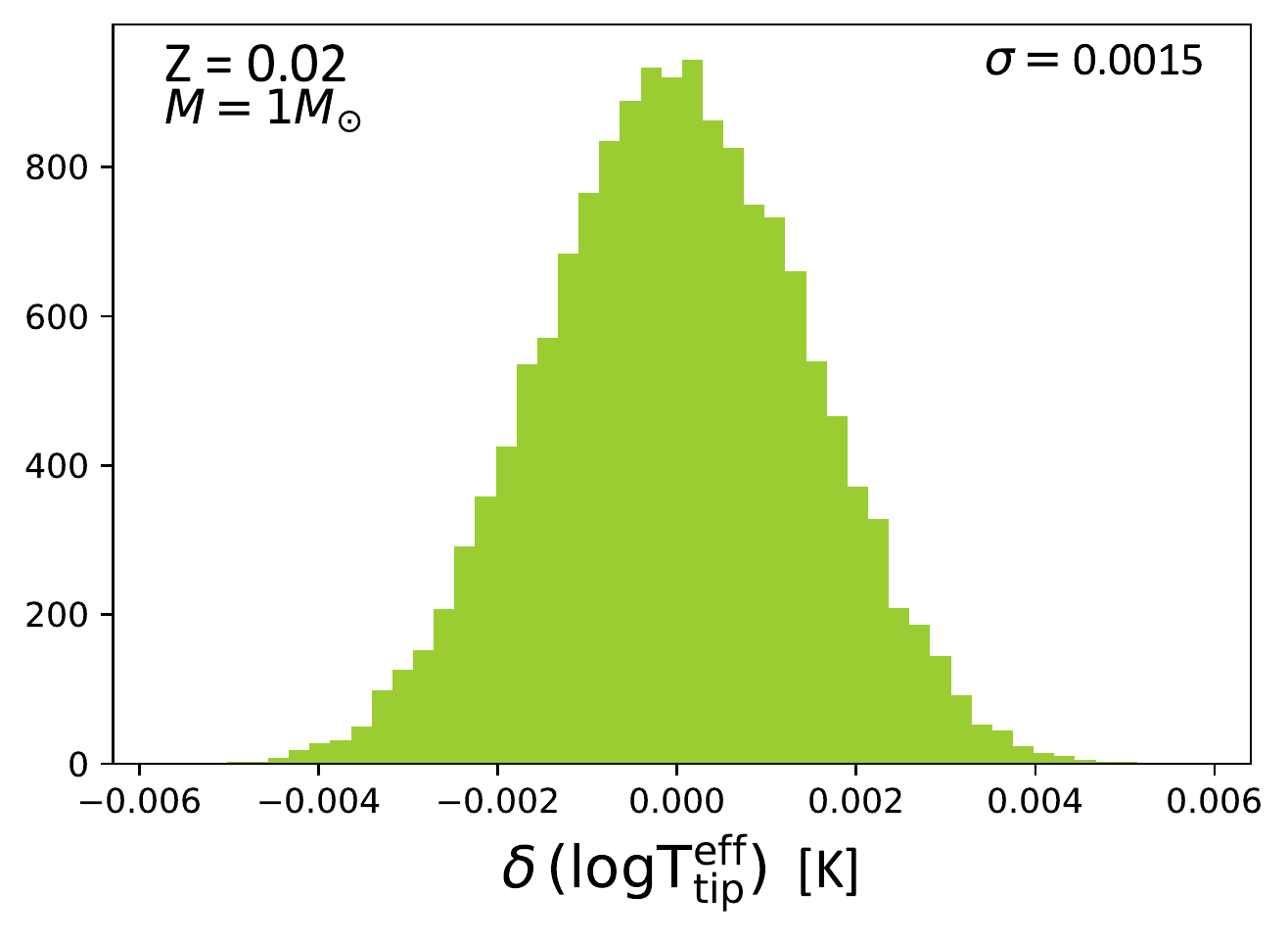}

\caption{Same as Figure \ref{fig:gauss_total}, but for the effective temperature at Tip luminosity.} \label{fig:gauss_total_Teff}
\end{figure*}

\section{TRGB luminosity from simultaneous variation of all input physics} \label{sec:full_analysis}
\begin{table*}
    \centering
    
    \caption{ Variation of the Tip luminosity ($\delta \log L/$\lsun) due to the most dominant uncertainties of input physics for different choices of the initial mass. The flag ``linear" corresponds to the linear addition of the uncertainties coming from all the quantities labelled with "1" in Table~\ref{tab:pert_eff}, according to equation (\ref{eq:sigma_linear}). The flag ``full" corresponds to the uncertainty computed through the simultaneous (Monte-Carlo) variation of all quantities labelled with "1" in Table~\ref{tab:pert_eff} except for helium (``$Y$") and boundary conditions (``BC"). Some histograms for the Monte-Carlo analysis are shown in Figure \ref{fig:gauss_total_L}. It is evident that the linear addition of individual uncertainties is in very good agreement with the Monte-Carlo analysis for most of the cases, with a mean difference between the two at $1.5 \%$. The ``full" (Monte-Carlo) case is based on about $15,000$ samples. }
    \begin{tabular}{r|rrrrrrr}
    $Z$ & $M=0.80$ & $M=0.90$ & $M=1.00$ & $M=1.10$ & $M=1.20$ & $M=1.30$ & $M=1.40$\\
    \hline
    0.0001 (linear) & 0.0142 & 0.0144 & 0.0148 & 0.0154 & 0.0166 & 0.0190 & 0.0246\\
    0.0006 (linear) & 0.0148 & 0.0139 & 0.0147 & 0.0146 & 0.0156 & 0.0170 & 0.0165\\
    0.0010 (linear) & 0.0147 & 0.0146 & 0.0149 & 0.0150 & 0.0151 & 0.0159 & 0.0165\\
    0.0060 (linear) & 0.0149 & 0.0152 & 0.0151 & 0.0153 & 0.0153 & 0.0156 & 0.0156\\
    0.0100 (linear)& 0.0150 & 0.0152 & 0.0152 & 0.0153 & 0.0155 & 0.0152 & 0.0156\\
    0.0130 (linear) & 0.0153 & 0.0154 & 0.0154 & 0.0157 & 0.0157 & 0.0159 &  0.0161\\
    0.0160 (linear) & 0.0152 & 0.0154 & 0.0155 & 0.0157 & 0.0158 & 0.0159 &  0.0159\\
   0.0200 (linear) & 0.0154 & 0.0156 & 0.0157 & 0.0158 & 0.0159 & 0.0162 &  0.0160\\
    \hline \hline 
    0.01 (full)  &  0.0149 &  0.0152 & 0.0152 & 0.0151 & 0.0156 & 0.0152 & 0.0157 \\
    0.02 (full)  &  0.0152 &  0.0154 &  0.0155 & 0.0151 & 0.0156 & 0.0159 & 0.0156 \\
    \hline
    \end{tabular}
    \label{tab:ind_tot}
\end{table*}
The goal of this section is to predict the variation on the TRGB luminosity from the {\it simultaneous} perturbations of all input physics, as opposed to the previous analysis of individual contributions. We will show that the results are in good agreement with the linear addition of each individual perturbation, providing robust justification for the simpler, linear addition of uncertainties. 

To predict the TRGB luminosity we randomly sample sets of values for (\krad, \kcon, $3\alpha$, $^{14}$N(p,$\gamma$)$^{15}$O, $\nu$) and compute the luminosity for each of them, with total of about $15,000$ iterations. Each of the five parameters is assumed to be a random variable sampled from a Gaussian distribution with a mean value defined by our reference models and a respective $\sigma$ value according to Table $\ref{tab:pert_eff}$ for two indicative metallicities and three initial masses. The results are shown in Figure \ref{fig:gauss_total_L}. The uncertainty on the Tip luminosity increases with metallicity and initial mass. For the maximum metallicity in our sample, and for initial mass of $1 M_{\odot}$, we find 
$
\delta \left( \log L_{\rm{Tip}} \right) = 0.0154. \label{delta_L_max}
$
As was shown in the analysis of individual uncertainties, the main contribution to the predicted Tip luminosity comes from the uncertainty in the radiative opacity.  We also compared the predicted total luminosity uncertainty at the Tip with that obtained by simply adding the individual contribution. In this latter case, under the hypothesis of 1) a linear response of the model to each perturbations, and 2) Gaussianly distributed parameters, the total $\sigma_\mathrm{tot}$ can be evaluated as the squared sum of each individual $\sigma_j$ as
\begin{equation}
    \sigma_\mathrm{tot} = \sqrt{\sum_{j=1}^{n_p} \sigma_j^2}. \label{eq:sigma_linear}
\end{equation}

A detailed comparison is shown in Table \ref{tab:ind_tot}. For example, for the maximum metallicity in our sample, $Z=0.020$, and $M=1$\msun{}, a comparison between the variation of the Tip luminosity computed according to the Monte-Carlo analysis and the linear addition of individual uncertainties (equation (\ref{eq:sigma_linear})) respectively, shows a difference by less than $2\%$. It turns out that, the simpler, linear combination of individual uncertainties can be safely adopted to estimate the total uncertainty on Tip luminosity.

\begin{figure}
\centering
\includegraphics[width=0.98\columnwidth]{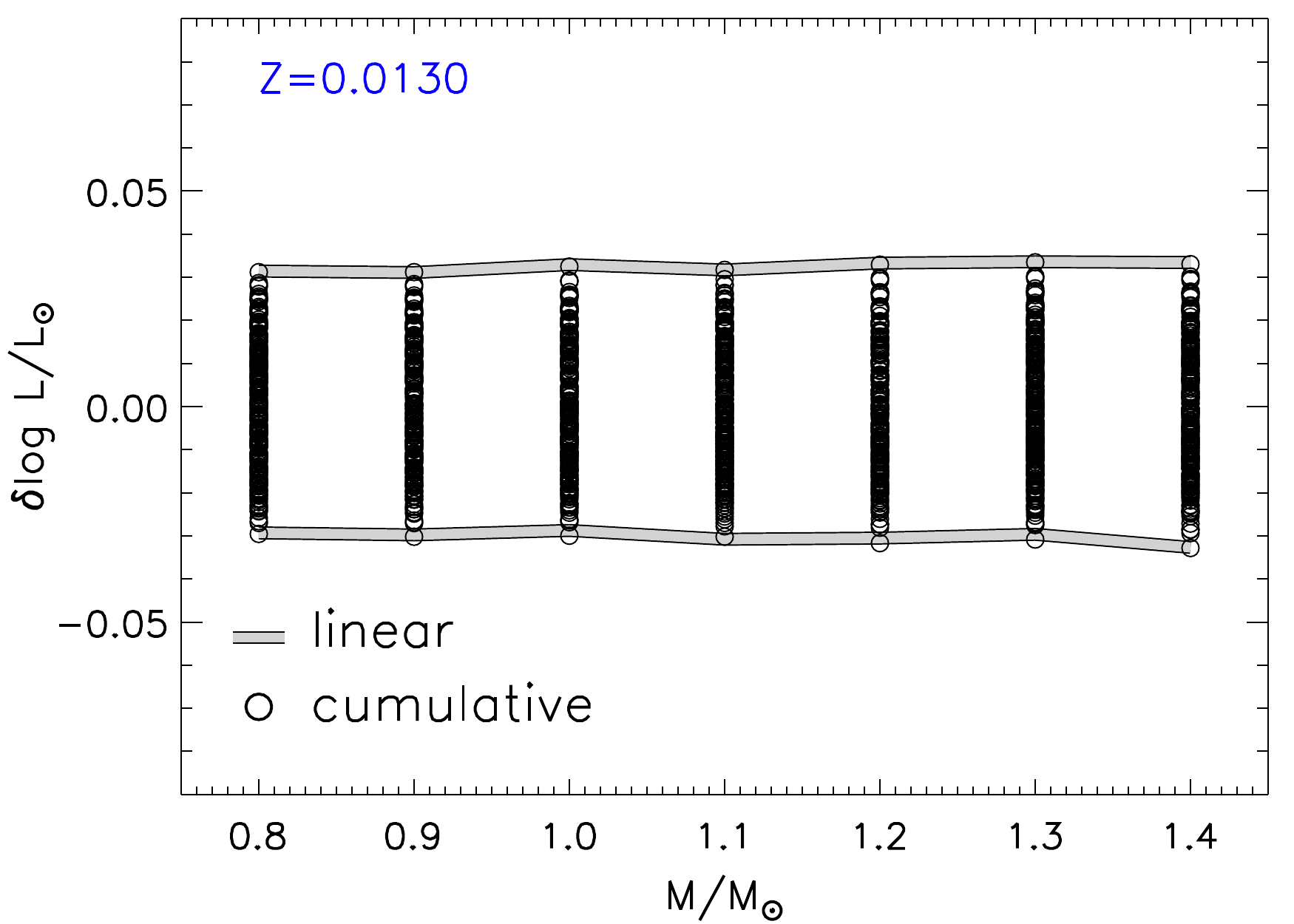}
\caption{Comparison between the TRGB luminosity variation obtained by linearly adding the individual perturbations (solid grey line) and the direct computations of models where the parameters are perturbed simultaneously (empty circles), for the quoted values of the stellar mass at solar chemical composition. The models account for the perturbation of \krad, \kcon, $^{14}$N(p,$\gamma$)$^{15}$O, $\alpha(\alpha\alpha,\gamma)^{12}$C and $\nu$. The top panel considers a metallicity $Z = 0.01$ and the lower panel $Z = 0.001$, which are the highest and lowest metallicity values of our analysis. }
\label{fig:linear_cumulative}
\end{figure}

As an additional test, Figure~\ref{fig:linear_cumulative} shows the maximum luminosity variation at the Tip, due to the simultaneous perturbation of the relevant quantities. This variation is different from that obtained using a Monte-Carlo analysis, since in this case, we consider only the global variation due to the perturbation of each quantity at $1\sigma$. Thus, this case is more similar to a linear combination of individual luminosity variations. To test the hypothesis of linearity, in this case, we simply add the luminosity variation due to individual perturbations to reproduce the edges of the full computation case. Figure~\ref{fig:linear_cumulative} shows that the linear model is capable of reproducing with very good approximation the edges obtained in the full computation case.

Whereas the Tip luminosity is the key observable for the construction of the distance ladder, the total radius and effective temperature of the star at the TRGB play an important role in asteroseismic studies. Therefore, we repeated the previously described analysis for both radius and effective temperature at Tip luminosity, with the results shown in Figures \ref{fig:gauss_total_R} and \ref{fig:gauss_total_Teff} respectively. Similarly to the luminosity, the uncertainty on \teff{} and radius can be well reproduced combining the individual perturbations.

The results discussed in this section give a robust support to the possibility of studying the uncertainties on the TRGB model producing models with single-perturbed parameter/input physics, which can significantly simplify computations. In particular, the number of models required to evaluate the uncertainty coming form $n$-parameters scales as $2\times n+1$ (under the inclusion of models for only $-1,0,1\sigma$), which is to be compared with the cumulative case where the number of models scales as $3^n$.

\begin{figure}
\centering
\includegraphics[width=0.98\columnwidth]{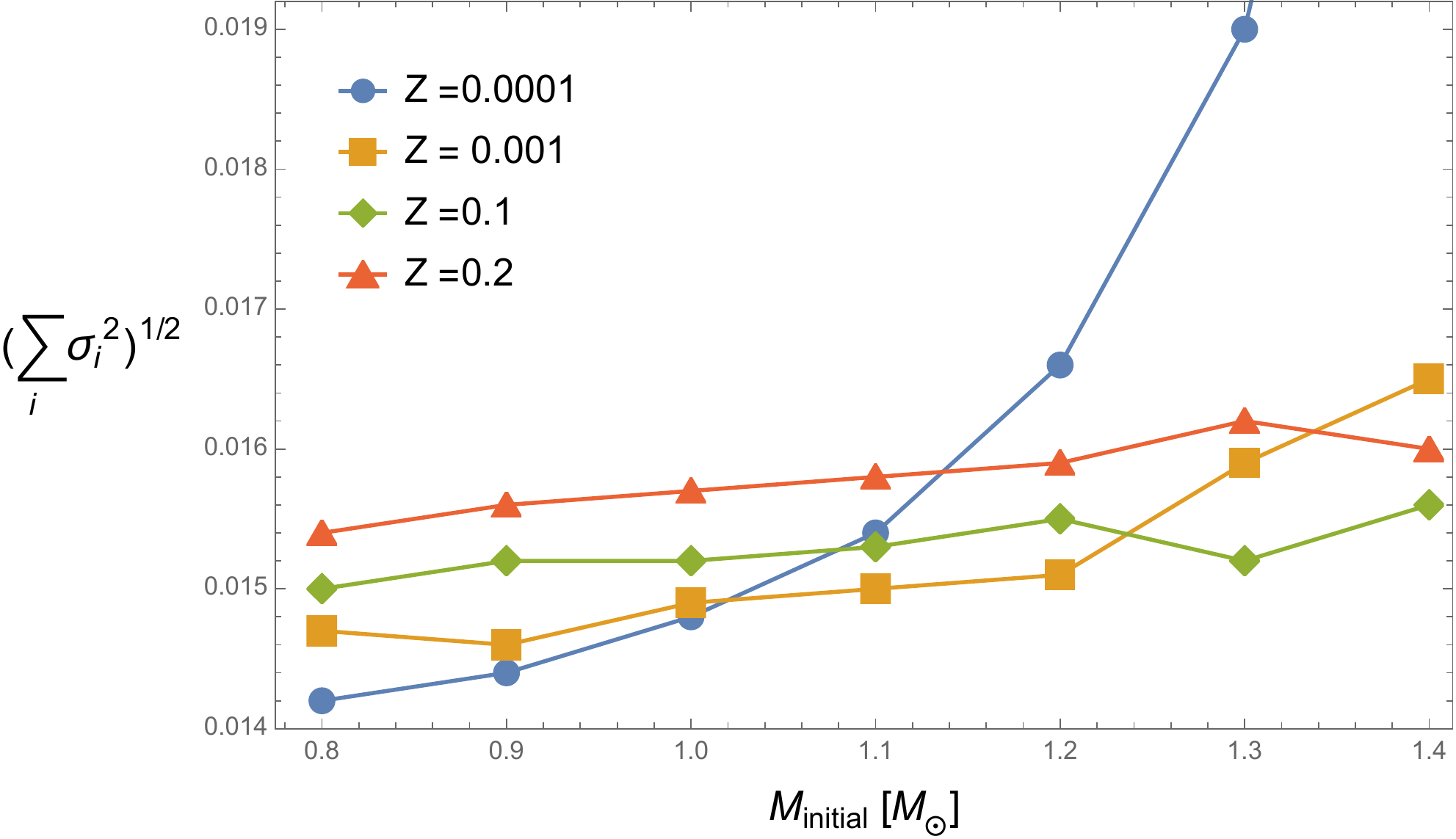}
\caption{The combined variation of the Tip luminosity computed by linearly adding all individual contributions to it coming from the quantities flagged with ``1" in Table~\ref{tab:pert_eff}, and according to equation (\ref{eq:sigma_linear}). The values used are according to Table \ref{tab:ind_tot}. Each curve corresponds to different metallicity $Z$. The linear addition of errors is to be compared with the Monte-Carlo analysis where all input physics are perturbed simultaneously. }
\label{fig:errors_individual}
\end{figure}

\section{From luminosity to magnitudes} \label{sec:magnitudes} 
\begin{landscape}
\begin{table*}
    \centering
    \caption{Variation of the predicted Tip magnitudes for some values of the metallicity $Z$ and total mass $M$ (in \msun), in different photometric bands: 1) Johnson/Cousin ($B$,$V$,$Rc$,$Ic$), Gaia ($G$, $Bp$, $Rp$), 2MASS ($J$, $H$, $Ks$) and WFC3 ($F438W$, $F555W$, $F625W$, $F814W$, $F105W.IR$, $F110W.IR$, $F125W.IR$, $F140W.IR$, $F160W.IR$). More bands and metallicities are available online in the full table.}
    \begin{tabular}{l|rrrrrrrrrrr}
    & & \multicolumn{4}{c}{Johnson/cousin} & \multicolumn{3}{c}{Gaia} & \multicolumn{3}{c}{2MASS}\\
    $Z$ & $M$  & $B$ & $V$ & $Rc$ & $Ic$ & $G$ &  $Bp$ & $Rp$ & $J$ & $H$ & $Ks$\\
    \hline
    0.0001 &  0.80 &  0.373 &  0.186 &  0.102 &  0.048 &  0.108 &  0.200 &  0.054 &  0.073 &  0.134 &  0.147 \\
    0.0001 &  0.90 &  0.359 &  0.176 &  0.096 &  0.046 &  0.103 &  0.192 &  0.052 &  0.076 &  0.136 &  0.150 \\
    0.0001 &  1.00 &  0.342 &  0.164 &  0.089 &  0.045 &  0.097 &  0.182 &  0.049 &  0.079 &  0.137 &  0.150 \\
    0.0001 &  1.10 &  0.326 &  0.153 &  0.083 &  0.046 &  0.092 &  0.173 &  0.048 &  0.083 &  0.139 &  0.152 \\
    0.0001 &  1.20 &  0.309 &  0.142 &  0.077 &  0.047 &  0.086 &  0.162 &  0.048 &  0.087 &  0.141 &  0.154 \\
    0.0010 &  0.80 &  0.372 &  0.239 &  0.152 &  0.084 &  0.141 &  0.231 &  0.093 &  0.047 &  0.107 &  0.128 \\
    0.0010 &  0.90 &  0.361 &  0.228 &  0.144 &  0.080 &  0.135 &  0.222 &  0.088 &  0.050 &  0.111 &  0.132 \\
    0.0010 &  1.00 &  0.339 &  0.210 &  0.132 &  0.074 &  0.127 &  0.206 &  0.082 &  0.052 &  0.114 &  0.134 \\
    0.0010 &  1.10 &  0.317 &  0.193 &  0.121 &  0.069 &  0.118 &  0.192 &  0.076 &  0.054 &  0.118 &  0.136 \\
    0.0010 &  1.20 &  0.301 &  0.179 &  0.112 &  0.064 &  0.110 &  0.180 &  0.070 &  0.057 &  0.121 &  0.139 \\
    0.0130 &  0.80 &  0.544 &  0.636 &  0.459 &  0.368 &  0.350 &  0.586 &  0.306 &  0.055 &  0.066 &  0.069 \\
    0.0130 &  0.90 &  0.713 &  0.841 &  0.596 &  0.471 &  0.452 &  0.774 &  0.392 &  0.067 &  0.087 &  0.091 \\
    0.0130 &  1.00 &  0.837 &  1.024 &  0.687 &  0.490 &  0.499 &  0.950 &  0.419 &  0.067 &  0.095 &  0.103 \\
    0.0130 &  1.10 &  0.941 &  1.186 &  0.768 &  0.489 &  0.534 &  1.111 &  0.433 &  0.065 &  0.101 &  0.114 \\
    0.0130 &  1.20 &  0.964 &  1.243 &  0.798 &  0.437 &  0.528 &  1.178 &  0.405 &  0.068 &  0.109 &  0.126 \\
    \hline
    & & \multicolumn{10}{c}{Wide Field Camera (WFC3)}  \\
    $Z$ & $M$ & $F438W$ & $F555W$ & $F625W$ & $F814W$ & IR:$F105W$ & $F110W$ & $F125W$ & $F140W$ & $F160W$ &\\
    \hline
    0.0001 &  0.80 &  0.408 &  0.200 &  0.118 &  0.049 &  0.052 &  0.063 &  0.075 &  0.096 &  0.119 & \\
    0.0001 &  0.90 &  0.392 &  0.190 &  0.111 &  0.047 &  0.054 &  0.066 &  0.078 &  0.099 &  0.122 & \\
    0.0001 &  1.00 &  0.373 &  0.178 &  0.103 &  0.045 &  0.058 &  0.069 &  0.080 &  0.101 &  0.123 & \\
    0.0001 &  1.10 &  0.355 &  0.167 &  0.096 &  0.046 &  0.062 &  0.073 &  0.084 &  0.104 &  0.125 & \\
    0.0001 &  1.20 &  0.335 &  0.156 &  0.088 &  0.047 &  0.066 &  0.077 &  0.088 &  0.107 &  0.128 & \\
    0.0010 &  0.80 &  0.401 &  0.245 &  0.170 &  0.085 &  0.044 &  0.043 &  0.048 &  0.065 &  0.088 & \\
    0.0010 &  0.90 &  0.390 &  0.235 &  0.161 &  0.080 &  0.043 &  0.044 &  0.051 &  0.068 &  0.092 & \\
    0.0010 &  1.00 &  0.368 &  0.217 &  0.147 &  0.075 &  0.043 &  0.046 &  0.054 &  0.072 &  0.096 & \\
    0.0010 &  1.10 &  0.344 &  0.200 &  0.135 &  0.069 &  0.043 &  0.047 &  0.056 &  0.075 &  0.099 & \\
    0.0010 &  1.20 &  0.327 &  0.186 &  0.124 &  0.064 &  0.043 &  0.049 &  0.059 &  0.079 &  0.103 & \\
    0.0130 &  0.80 &  0.547 &  0.637 &  0.527 &  0.328 &  0.059 &  0.044 &  0.049 &  0.045 &  0.049 & \\
    0.0130 &  0.90 &  0.714 &  0.843 &  0.696 &  0.420 &  0.061 &  0.045 &  0.058 &  0.052 &  0.062 & \\
    0.0130 &  1.00 &  0.817 &  1.027 &  0.869 &  0.439 &  0.058 &  0.044 &  0.061 &  0.060 &  0.073 & \\
    0.0130 &  1.10 &  0.900 &  1.189 &  1.031 &  0.443 &  0.055 &  0.045 &  0.062 &  0.069 &  0.083 & \\
    0.0130 &  1.20 &  0.908 &  1.245 &  1.109 &  0.401 &  0.049 &  0.048 &  0.067 &  0.078 &  0.094 & \\
    \hline
    \end{tabular}
    \label{tab:mags}
\end{table*}
\end{landscape}
As a final step, we translate the previous results to magnitudes, as they are what is directly observed. To derive the magnitude of a stellar model in a specific photometric filter, we adopted the method described in \citet{girardi02}. The absolute magnitude $M_\lambda$ is defined as,
\begin{equation}
M_\lambda = -2.5 \log L/\mathrm{L}_{\sun} -  bc_\lambda(T_\mathrm{eff},g,\mathrm{[Fe/H]})
\label{eq:mag}
\end{equation}
where $L$, \teff, $g$ and [Fe/H] are the luminosity, effective temperature, surface gravity and [Fe/H] of the star and $bc_\lambda($\teff,$g$,[Fe/H]) are the bolometric corrections obtained using the MARCS  spectra library \citep{gustafsson08}\footnote{\url{https://marcs.astro.uu.se/}}. Thus, the uncertainty on the magnitude has been evaluated considering pairs of reference  (M$_\lambda$(ref)) and perturbed (M$_\lambda$(pert)$|_j$) models respectively, 
\begin{equation}
\delta M_\lambda = \frac{1}{\sqrt{N}}\sqrt{\sum_j (M_\lambda(\mathrm{pert})|_j - M_\lambda(\mathrm{ref}))^2}.
\end{equation}
The variation of the magnitude has two contributions - one from the luminosity variation (bolometric magnitude), and another from the variation of the bolometric corrections due to $\delta $\teff{} and $\delta g$. In this regard, among the quantities listed in Table~\ref{tab:pert} there are some that affect mainly the Tip luminosity (e.g \kcon, reaction rates, neutrinos), others that affect mainly \teff{} and $g$ (e.g BCs, \ml), and others that affect $L$, \teff{} and $g$ (e.g \krad and Y). Thus, taking into account only the variation on the Tip luminosity to derive the corresponding variation in Tip magnitude, as sometimes done in the literature, can lead to a wrong estimation of the actual magnitude uncertainty. To avoid this problem, we evaluated for each perturbed model the magnitude using eq.~(\ref{eq:mag}) with the proper values of $L$, \teff{} and $g$. We considered only quantities for which we can define a meaningful value of $\sigma$, neglecting $\beta_\mathrm{ov}$ and electron screening. The effect of mass loss will be discussed later.

We performed this analysis for all the masses and metallicities in our sample, considering popular bands used in the literature, namely $B$, $V$, $Rc$, $Ic$ (Johnson/Cousins), $J$, $H$, $Ks$ (2MASS), $G$, $Bp$, $Rp$ (Gaia), $F218W$, $F225W$, $F275W$, $F336W$, $F390W$, $F438W$, $F475W$, $F555W$, $F606W$, $F625W$, $F775W$, $F814W$, $F105W.IR$, $F110W.IR$, $F125W.IR$, $F140W.IR$, $F160W.IR$ (WFC3). The estimated variation of the magnitude for some bands and masses are shown in Table~\ref{tab:mags}. As a general trend, optical bands (e.g. $B$, $V$ or equivalent filters in Gaia or WFC3) are largely affected by the uncertainties in input physics and exhibit a strong dependence with the metallicity. Magnitudes in optical bands can vary between 0.1-0.2~mag at $Z=0.0001$ up to 1.2~mag at solar metallicity. Near infrared bands (NIR) are less dependent on the input physics/parameters adopted in the model. For the $Ic$ band (or equivalently, the $F814W$ WFC3 band) we obtained a variation of about 0.05~mag ($Z=0.0001$) which increases to 0.08~mag at $Z=0.0010$ and 0.49~mag at $Z=0.013$ respectively. Increasing the filter wavelength, especially in the cases of $\lambda_\mathrm{eq} \ga 1000\mu$m (e.g. 2MASS), the magnitude variation  ranges between 0.05-0.1~mag almost independently on metallicity. 

We also checked the impact of the mass loss on the Tip magnitude. We recall that the mass loss has a small impact on the Tip luminosity, but it affects the effective temperature. Thus, the effect on the Tip magnitude of the adopted mass loss comes mainly from the variation of \teff. We show in Table~\ref{tab:eta_mag} the magnitudes change obtained passing from a standard value of $\eta_\mathrm{Reim.}=0.3$ to 0. As in the previous cases, the effect increases with the metallicity and it can be quite large in the optical bands (up to 0.6-0.7~mag at $Z=0.0130$ for the $B$, $V$ bands), whereas it is smaller in the IR, especially in the $J$ band where the magnitude variation ranges approximately between 0.010-0.030~mag. 
\begin{table*}
    \centering
    \caption{Magnitude variation due to the mass loss: differences between the reference set of models ($\eta_\mathrm{Reim.}=0.3$) and the models without mass loss. We listed only Johnson/cousins ($B$, $V$, $Rc$, $Ic$) and 2MASS ($J$, $H$, $Ks$) filters.}
    \begin{tabular}{lrrrrrrrr}
    $Z$ & $M$ & $B$ & $V$ & $Rc$ & $Ic$ & $J$ & $H$ & $Ks$\\
    \hline
    0.0001 &  0.80 &  0.053 &  0.026 &  0.013 &  0.003 &  0.006 &  0.011 &  0.013 \\
    0.0001 &  0.90 &  0.051 &  0.024 &  0.010 &  0.001 &  0.010 &  0.017 &  0.019  \\
    0.0001 &  1.00 &  0.043 &  0.021 &  0.011 &  0.004 &  0.005 &  0.011 &  0.012 \\
    0.0010 &  0.80 &  0.126 &  0.072 &  0.043 &  0.019 &  0.006 &  0.021 &  0.026  \\
    0.0010 &  0.90 &  0.121 &  0.069 &  0.040 &  0.017 &  0.010 &  0.028 &  0.034 \\
    0.0010 &  1.00 &  0.085 &  0.046 &  0.026 &  0.010 &  0.010 &  0.024 &  0.029  \\
    0.0130 &  0.80 &  0.190 &  0.174 &  0.131 &  0.106 &  0.007 &  0.003 &  0.006  \\
    0.0130 &  0.90 &  0.479 &  0.546 &  0.374 &  0.284 &  0.027 &  0.038 &  0.043  \\
    0.0130 &  1.00 &  0.558 &  0.692 &  0.433 &  0.293 &  0.023 &  0.043 &  0.053 \\
    \hline
    \end{tabular}
    \label{tab:eta_mag}
\end{table*}

Another contribution to the error budget of the Tip magnitude is the uncertainty on the bolometric corrections obtained from synthetic spectra. Unfortunately, the uncertainty on the synthetic spectral energy distribution is not available, so, the best we could do was to roughly estimate the impact on the predicted magnitudes of adopting different spectra libraries. To do this, we computed an additional set of bolometric corrections using the \citet{allard11} spectra, which cover the whole temperature/gravity/[Fe/H] range required for our analysis and  we compared the Tip magnitudes in different photometric bands with those obtained using our reference set of bolometric corrections. We found a relatively large variation of the predicted Tip magnitudes, which increases with metallicity. In the Johnson/Cousins bands we found the following variations: 0.14-0.24~mag ($Z=0.0001$), 0.18-0.30~mag ($Z=0.001$) and 0.20-0.45~mag ($Z=0.013$). For 2MASS bands the effect is larger, ranging between 0.20-0.30~mag ($Z=0.0001$), 0.25-0.40~mag ($Z=0.001$) and 0.35-0.60~mag ($Z=0.013$). Our results are consistent with those of \citet{viaux13} and \cite{serenelli17}, that is, the systematic error from the adoption of a particular spectral library is one of the main error sources that affects the the Tip magnitude.

\section{Summary of results} \label{sec:summary}
Understanding the modelling and theoretical uncertainties of the Tip luminosity of TRGB stars is crucial for the the inference of distances in astrophysics and the accurate measurement of the local value of the Hubble parameter. Here, we presented a detailed investigation of the theoretical uncertainties on the TRGB luminosity based on a new grid of models evolved from pre-main sequence up to the TRGB stage. To our best of knowledge, this is the most complete analysis in the literature to date. In summary:
\begin{itemize}
\item[] 1. We presented a new generation of calibrated TRGB models, evolved from pre-main sequence up to the helium ignition stage and for a metallicity range $Z = 0.001-0.02$. The list of input physics parameters and their effect on the Tip luminosity is found in Tables \ref{tab:pert} and \ref{tab:pert_eff}, while a detailed discussion on each of them in Section \ref{sec:uncertainties}. The accuracy of our numerical method is discussed in Section \ref{sec:timestep} (see also Figure \ref{fig:time_step2}), while a comparison with other codes is shown in Figure \ref{fig:comp}.  
\\

\item[] 2. We showed that the combined effect of the most dominant uncertainties considered leaves a residual uncertainty of $1.6 \%$ on the TRGB luminosity, with the biggest contribution coming from the modelling of the radiative opacity. The results of our Monte-Carlo analysis are shown in Figures \ref{fig:gauss_total_L}, \ref{fig:gauss_total_R} and \ref{fig:gauss_total_Teff}, while those for the individual effect of uncertainties in Figure \ref{fig:lin_fit} and Table \ref{tab:fit} respectively. Table \ref{tab:ind_tot} provides an overview of all computed uncertainties on the Tip luminosity. The corresponding uncertainties in the photometric magnitudes are listed in Table \ref{tab:mags}, and discussed in Section \ref{sec:magnitudes}. The biggest uncertainty in Tip magnitudes comes from the choice of spectral library used.
\\

\item[] 3. We demonstrated that, the uncertainty on the TRGB luminosity found by linearly adding individual uncertainties is consistent with the one based on the simultaneous (Monte-Carlo) variation of the relevant uncertainties. This result strengthens the previously found linear response of TRGB models in the literature, and can significantly simplify future computations. A comparison between the linear with the full analysis is shown in Figure \ref{fig:linear_cumulative} and discussed in Sections \ref{sec:linearity} and \ref{sec:full_analysis}.
\end{itemize}

We believe our results we will greatly contribute to the improvement of modelling of the TRGB, and the accurate inference of the local Hubble parameter, in light of future astrometric data from Gaia and other surveys. It would be interesting to investigate how the current modelling can be further improved with asteroseismic observations. For that, one would need an investigation of the effect of input physics on the pulsation frequencies. Finally, we note that our grid of models is available upon reasonable requests to aid future analyses on the calibration of the TRGB luminosity.

\section*{Acknowledgements}
Emanuele Tognelli is supported by the Czech Grant Agency (GAČR), under the grant number 21-16583M. Ippocratis D. Saltas acknowledges support from the Czech Academy of Sciences under the project number LQ100102101.

\bibliographystyle{mn2e}
\bibliography{bibliography}


\label{lastpage}

\end{document}